\documentclass[11pt, twoside]{article}
\usepackage{amsfonts}
\usepackage{amsmath}
\usepackage{amssymb}
\usepackage{amsthm}
\usepackage{latexsym}
\usepackage{natbib}
\usepackage{graphicx}
\usepackage{xcolor}
\usepackage{fancyhdr}
\usepackage{fancybox}
\usepackage[T1]{fontenc}
\usepackage[scaled]{helvet}

\title{A plug--in rule for bandwidth selection in\\ circular density estimation}
\author{M. Oliveira*, R. M. Crujeiras and A. Rodr\'iguez--Casal\\ \small{Department of Statistics and Operations Research}\\ \small{University of Santiago de Compostela (Spain)}}
\date{}
\begin{document}
\maketitle

\begin{center}
\textbf{Abstract}
\end{center}
\begin{quotation}
A new plug--in rule procedure for bandwidth selection in kernel circular density estimation is introduced. The performance of this proposal is checked throughout a simulation study considering a variety of circular distributions exhibiting multimodality, peakedness and/or skewness. The plug--in rule behaviour is also compared with other existing bandwidth selectors. The method is illustrated with two classical datasets of cross--beds layers and animal orientation.\\

\end{quotation}

\vspace*{7cm}
\begin{quotation}
\noindent
\textbf{*Corresponding author:}\\
Mar\'ia Oliveira\\
Department of Statistics and Operations Research\\
Faculty of Mathematics - University of Santiago de Compostela (Spain)\\
e--mail: maria.oliveira1@rai.usc.es
\end{quotation}

\newpage
\section{Introduction}
Circular data appear in a large variety of disciplines, such as ecology, metereology or environmental sciences (see Mardia and Jupp, 2000 for examples). The analysis of circular data has been approached from parametric and nonparametric perspectives, existing a broad statistical literature on parametric methods. The classical von Mises distribution may be used to fit some real data, such as the azimuth dataset in Section 4, but multimodal distributions appear naturally in practical situations. A more flexible model considering mixtures of von Mises distributions to fit sudden infant deaths is used by Mooney et al. (2003), although the results did not suggest a multimodal situation. A parametric family for circular models with two diametrically opposed modes is introduced by Abe and Pewsey (2011), who illustrate its behaviour in animal orientation studies. More complex but unimodal features can be described by the four--parameter collection of distribution families proposed by Jones and Pewsey (2011), namely the inverse Batschelet distributions, which accounts for skewness and peakedness (far from the nicely bell--shaped von Mises distribution). In these contexts, parameter estimation can be done by maximum likelihood and Akaike Information Criterion (AIC) allows for fit diagnostic. However, in a general setting, determining a specific parametric family accounting for multimodality and/or asymmetry may not be an easy task.

An alternative to parametric families, both inferentially and as a descriptive tool, is the kernel density estimation proposed for the general case of spherical data by Hall et al. (1987), following the ideas of the classical kernel density estimator for linear data. Theoretical properties have been studied by Hall et al. (1987), who derived the expressions for bias, variance and loss (squared--loss and Kullback--Leibler discrepancy), and Klemel\"a (2000), who studied the estimation of the density derivatives. As in any nonparametric procedure, a smoothing parameter must be chosen, minimizing some error criterion. The use of cross--validation bandwidths is suggested by Hall et al. (1987) and more recently, Taylor (2008) derived a rule of thumb for bandwidth selection in circular density estimation. Although providing a simple choice in practice, the performance of this selector may be extremely poor in some distribution settings involving multimodality, peakedness or skewness. 

The goal of this work is to introduce a new procedure for selecting the smoothing parameter in kernel circular density estimation that performs well in distributional scenarios far from the von Mises case. Our proposal is based on the simple idea of the rule of thumb proposed by Taylor (2008), but allowing more flexibility in the underlying model. This further flexibility is achieved by considering mixtures of von Mises distributions as a reference for asymptotic mean integrated square error minimization.

This paper is organized as follows. Section 2 is devoted to the introduction of the kernel density estimator for circular data, revising different bandwidth selection procedures and introducing the new method. The performance of the new procedure is checked in a simulation study in Section 3, considering a wide class of circular density families, involving multimodality, peakedness and skewness. The technique is illustrated with some classical data in Section 4.

\section{Circular kernel density estimation}
\label{kernel_section}

Given a random sample of angles $\Theta_1,\Theta_2,\ldots,\Theta_n \in [0,2\pi)$ from some unknown density $f$, the kernel circular density estimator of $f$ is defined as:
\[
\hat f(\theta;\nu)=\frac{1}{n} \sum_{i=1}^{n} K_{\nu}(\theta-\Theta_i), \quad 0\leq \theta < 2\pi,
\]
where $K_{\nu}$ is the circular kernel function with concentration parameter $\nu>0$ (see Di Marzio et al., 2009). As a circular kernel, the von Mises distribution can be considered. Also known as the circular Normal, the von Mises distribution, $vM(\mu,\kappa)$, is a symmetric unimodal distribution characterized by a mean direction $\mu\in[0,2\pi)$, and concentration parameter $\kappa\geq 0$, with probability density function
\[
g(\theta;\mu,\kappa)=\frac{1}{2\pi I_0(\kappa)}\exp\left\{{\kappa \cos(\theta-\mu)}\right\}, \quad 0\leq \theta < 2\pi,
\]
where $I_r$ denotes the modified Bessel function of order $r$. With this specific kernel, the density estimator is given by:
\begin{equation}
\hat{f}(\theta; \nu)=  \frac{1}{n (2\pi) I_0(\nu)} \sum_{i=1}^{n} \exp{\left\{\nu \cos(\theta - \Theta_i)\right\}}, \quad 0\leq \theta < 2\pi.
\label{kernel_circular}
\end{equation}

A critical issue when using this estimator in practice is the choice of the smoothing parameter $\nu$. Large values of $\nu$ lead to highly variable (undersmoothed) estimators, whereas small values of $\nu$ imply low concentration of the kernel, providing oversmoothed estimators for the circular density. 

Usually, the bandwidth parameter is selected in order to minimize some error criterion, such as the mean integrated squared error (MISE, $MISE(\nu)=\mathbb E(\int(\hat f-f)^2)$). The asymptotic expression for the MISE (AMISE) is derived by Di Marzio et al. (2009). For the circular kernel estimator (\ref{kernel_circular}), the AMISE($\nu$) when $\nu\rightarrow\infty$ and $\sqrt{\nu}n^{-1}\rightarrow 0$ is given by:
\begin{equation}
AMISE(\nu)=\left\{ \frac{1}{16} \left[1-\frac{I_2(\nu)}{I_0(\nu)}\right]^2 \int_{0}^{2\pi} \left[f^{\prime\prime}(\theta)\right]^2 d\theta + \frac{I_0(2\nu)}{2n\pi \left(I_0(\nu)\right)^2}\right\},
\label{AMISE}
\end{equation}
where $f^{\prime\prime}$ denotes the second--order derivative of the target density to be estimated, which measures the curvature of $f$. Densities with marked modes will give a larger value of its integral, whereas the lowest value is achieved by a circular uniform model.  

A \emph{rule of thumb}, adapting the idea of Silverman (1986) for bandwidth selection in linear kernel estimation, was proposed by Taylor (2008).
Assuming that the data follow a von Mises distribution with concentration parameter $\kappa$, the bandwidth minimizing the AMISE can be estimated by
\begin{equation}
\hat{\nu}_{RT} = \left[\frac{3 n \hat{\kappa}^2 I_2(2\hat{\kappa})}{4 \pi^{1/2} I_0(\hat{\kappa})^2} \right]^{2/5},
\label{PI-selector}
\end{equation}
where $\hat\kappa$ is obtained by maximum likelihood. This selector performs satisfactorily in fitting unimodal symmetric distributions, without highly peaked modes but its behaviour can be dramatically misleading in the presence of antipodal modes and/or skewed distributions (see Section \ref{simulations}). A very simple example arises when mixing two population with opposite centers but in the same proportion and with the same concentration parameters. The estimate $\hat\kappa$ will return a value close to zero, which corresponds with a circular uniform distribution. Consequently, a small value for $\hat\nu_{RT}$ will be obtained resulting in an oversmoothed kernel estimator for the circular density. 

An alternative route would be to plug--in a more flexible distribution family as a reference density in the AMISE. For that purpose, a mixture of von Mises can be considered. A finite mixture of $M$ von Mises distributions, $vM(\mu_i, \kappa_i)$ with proportions $\alpha_i$, $i=1,\ldots,M$, has density:
\begin{equation}
g(\theta) = \sum_{i=1}^{M} \alpha_i  \frac{\exp\left\{\kappa_i \cos(\theta-\mu_i)\right\}}{2 \pi I_0(\kappa_i)}, \quad\mbox{with}\quad \sum_{i=1}^{M}{\alpha_i}=1. 
\label{vonMises_mixture}
\end{equation}
In fact, the circular kernel density estimator in (\ref{kernel_circular}) can be seen as a mixture of $n$ von Mises distributions, centered in the data sample and with common concentration parameter $\nu$.

The proposed plug--in bandwidth selector, $\hat\nu_{PI}$, is obtained as follows:
\begin{itemize}
\item[Step 1.] Based on the sample information, select the number of mixture components $M$ for the reference distribution.
\item[Step 2.] Estimate the parameters in the von Mises mixture (\ref{vonMises_mixture}), $(\mu_i,\kappa_i,\alpha_i)$, for $i=1,\ldots, M$ and compute the integral $\int(\hat f^{\prime\prime}(\theta))^2d\theta$. Plug--in this quantity in the AMISE expression (\ref{AMISE}) to get $\widehat{\mbox{AMISE}}(\nu)$.
\item[Step 3.] Minimize $\widehat{\mbox{AMISE}}(\nu)$ and obtain $\hat\nu_{PI}$.
\end{itemize}

For Step 1, the selection of the number of mixture components in the reference distribution can be done by AIC, considering different numbers of mixtures. Maximum likelihood estimation via EM algorithm is used for Step 2 (see Banerjee et al., 2005). The integral in Step 2 can be efficiently computed numerically, by quadrature methods. In Step 3, an optimization method can be used, in order to minimize the AMISE. Details on the algorithm implementation with further explanation referring to each step will be given in Section 3.

This type of plug--in rules are not the only alternative to smoothing parameter selection, and some other data--driven procedures were already proposed by Hall et al. (1987) using cross--validation ideas. In order to check the performance of the proposed method, its behaviour will be compared also with a likelihood cross--validation bandwidth. Precisely, the likelihood cross--validation bandwidth $\hat\nu_{LCV}$ is obtained by maximizing:
\begin{equation}
LCV(\nu)=\prod_{i=1}^{n}\hat{f}_{-i}(\theta_i;\nu),
\label{LCV}
\end{equation}
where $\hat{f}_{-i}$ denotes the circular kernel density estimator (\ref{kernel_circular}) leaving out the $i$-th observation. Our empirical experiments show a more stable behaviour of this selector compared with the classical least--squares cross--validation method (see also Taylor, 2008).

\begin{figure}
\begin{center}
\includegraphics[scale=0.22]{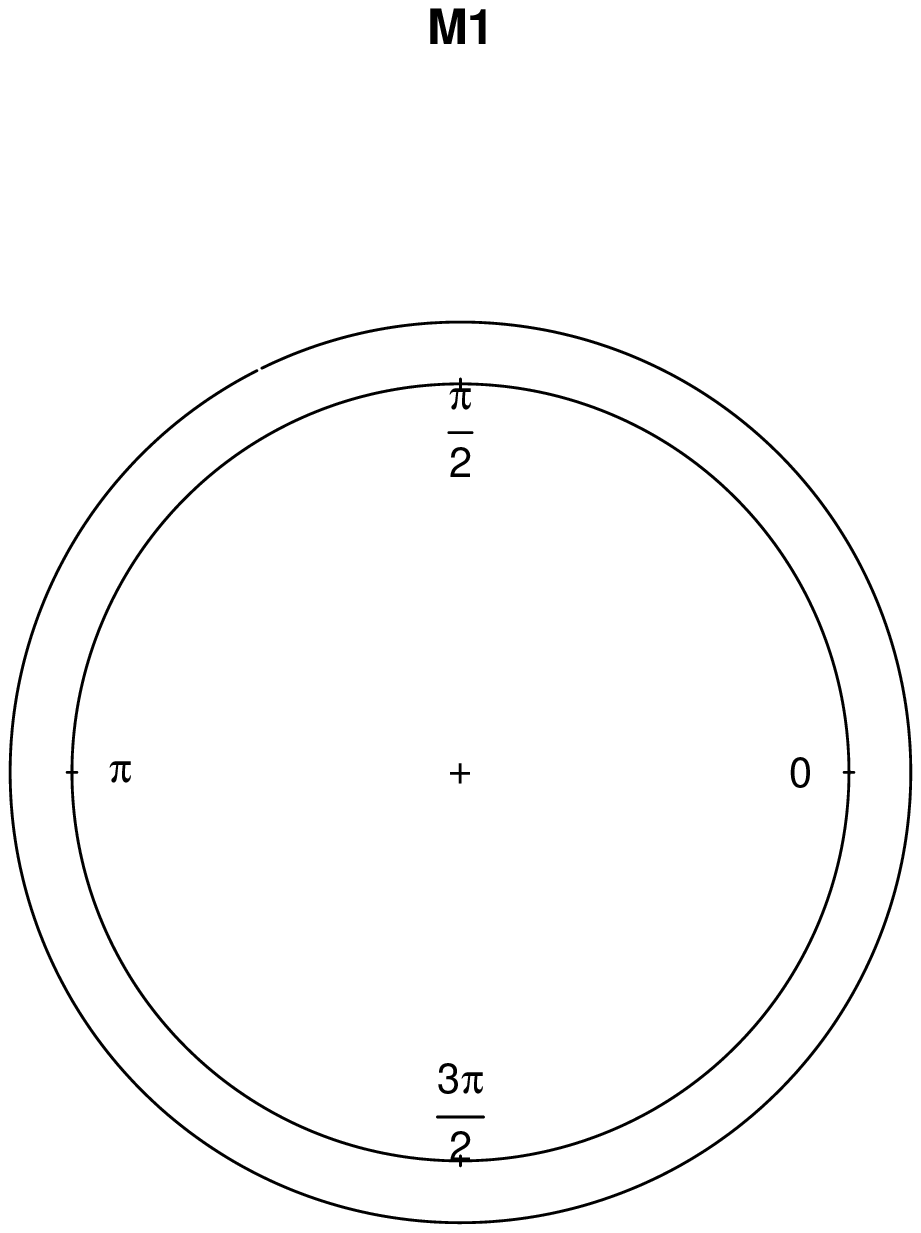}
\includegraphics[scale=0.22]{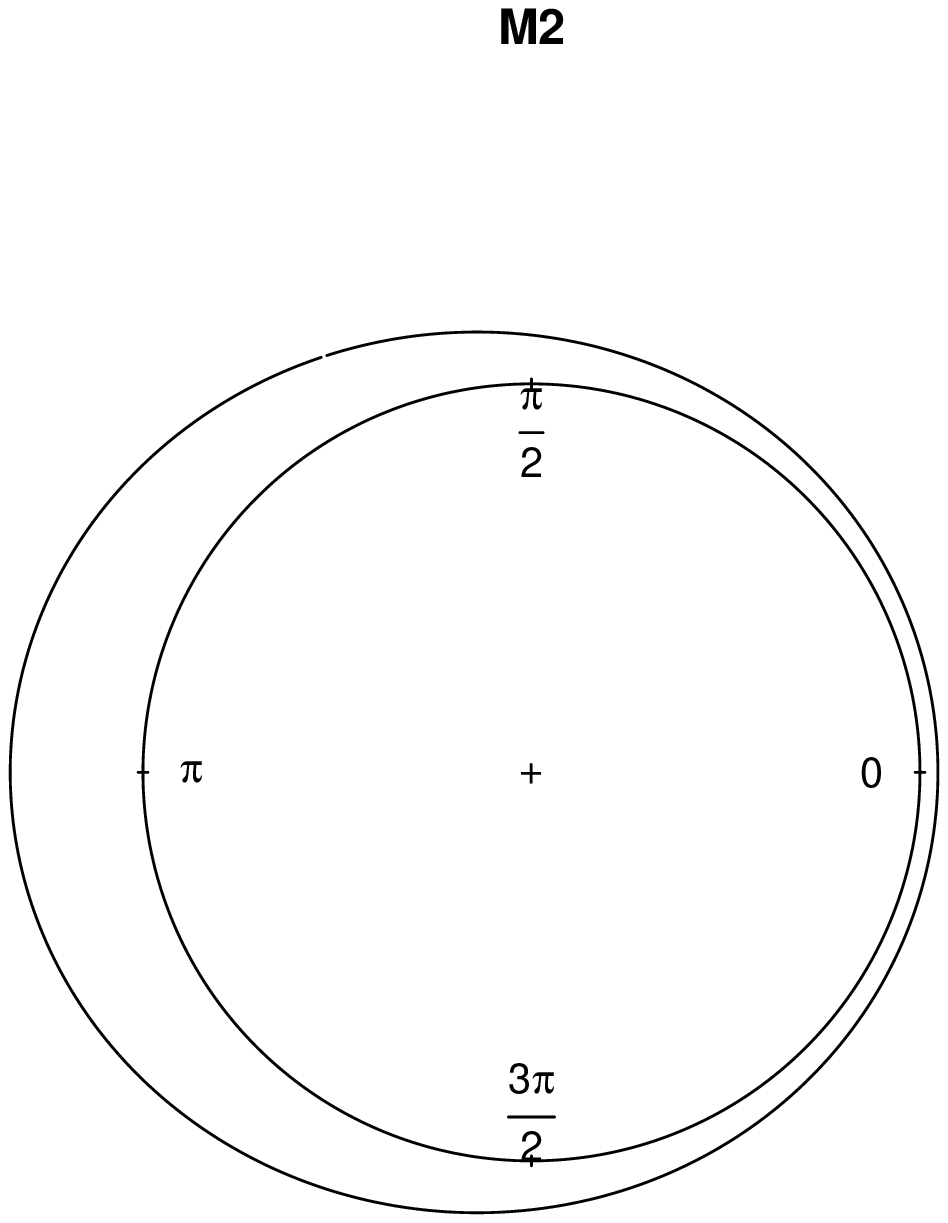}
\includegraphics[scale=0.22]{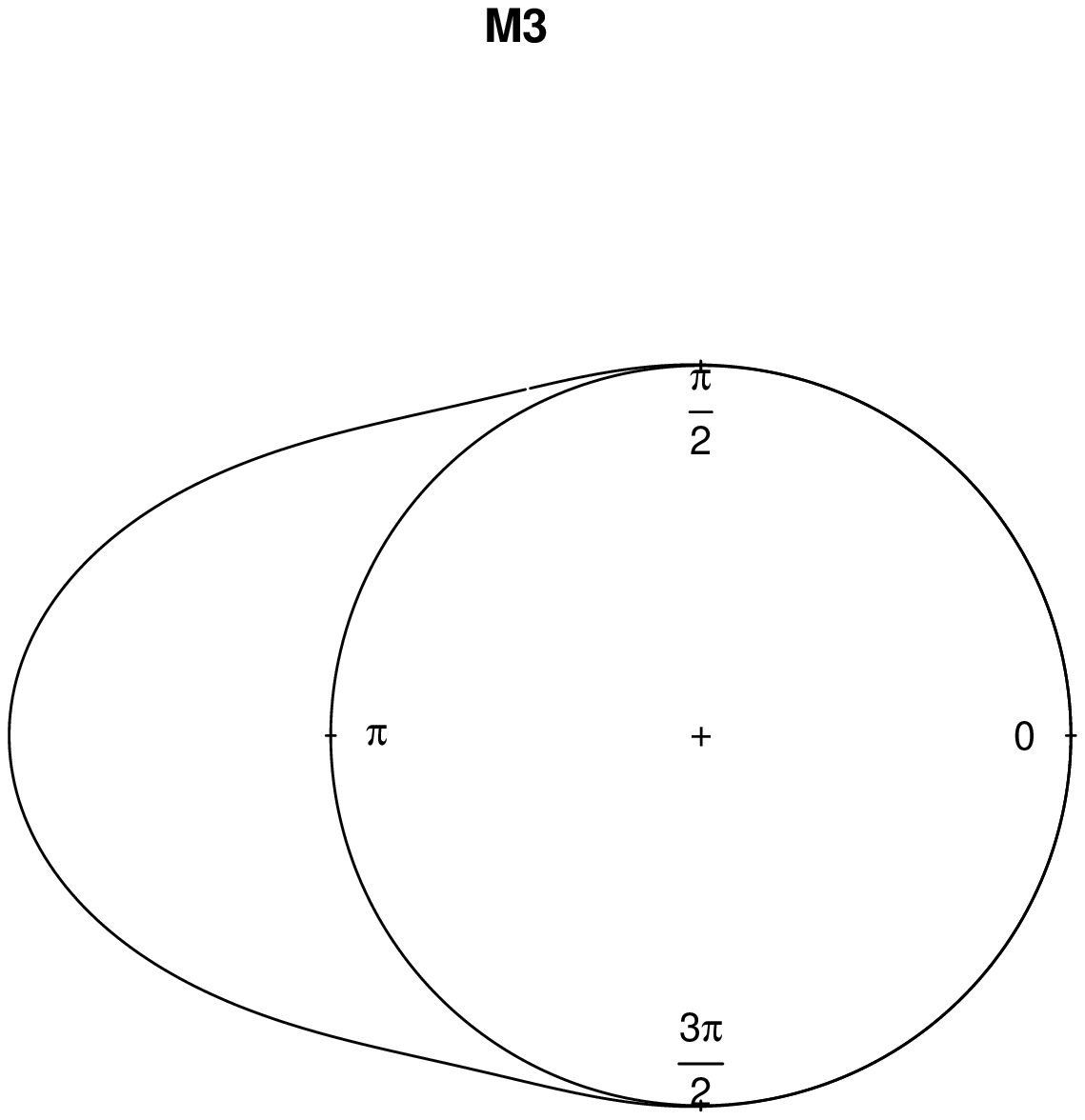}
\includegraphics[scale=0.22]{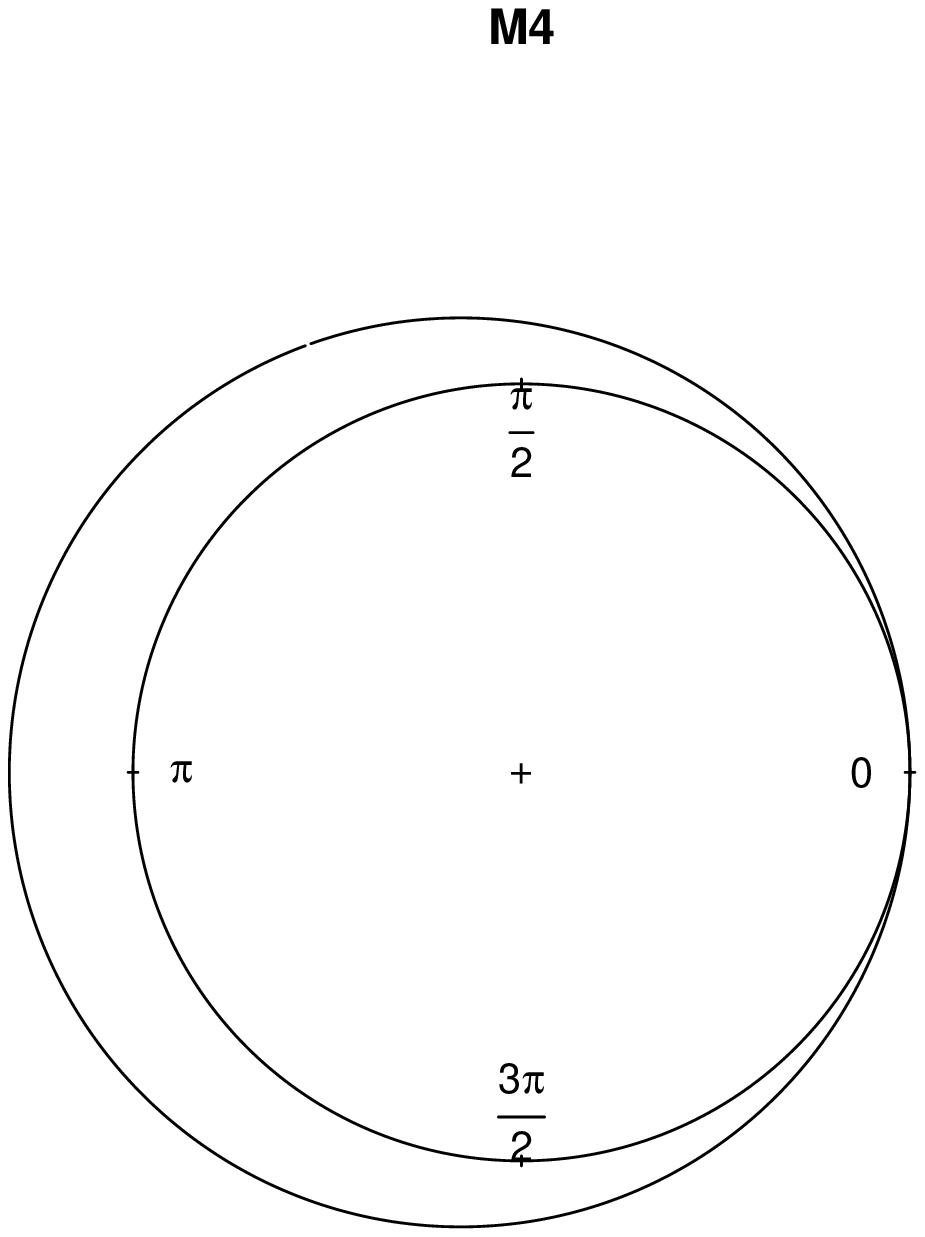}\\
\includegraphics[scale=0.22]{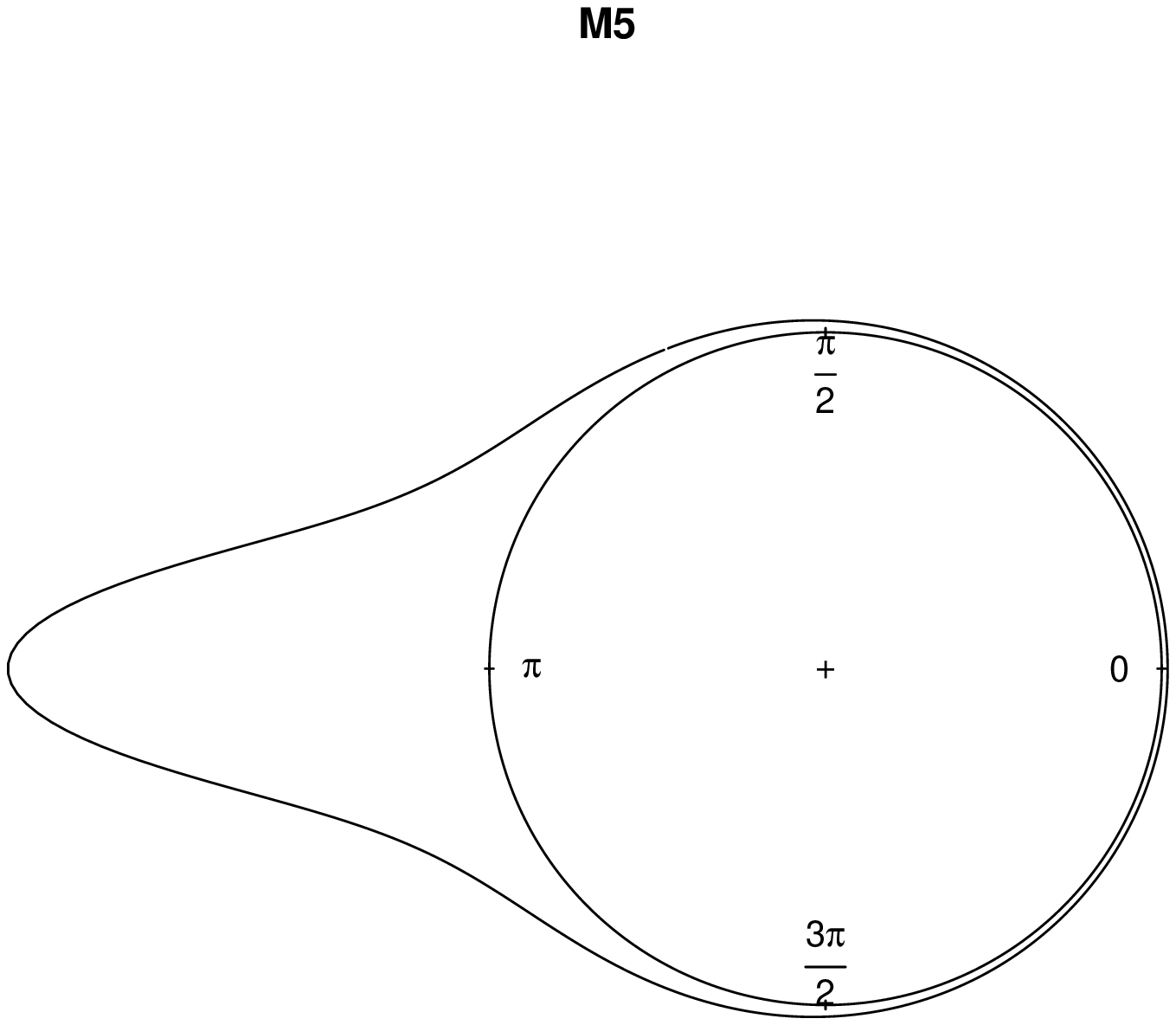}
\includegraphics[scale=0.22]{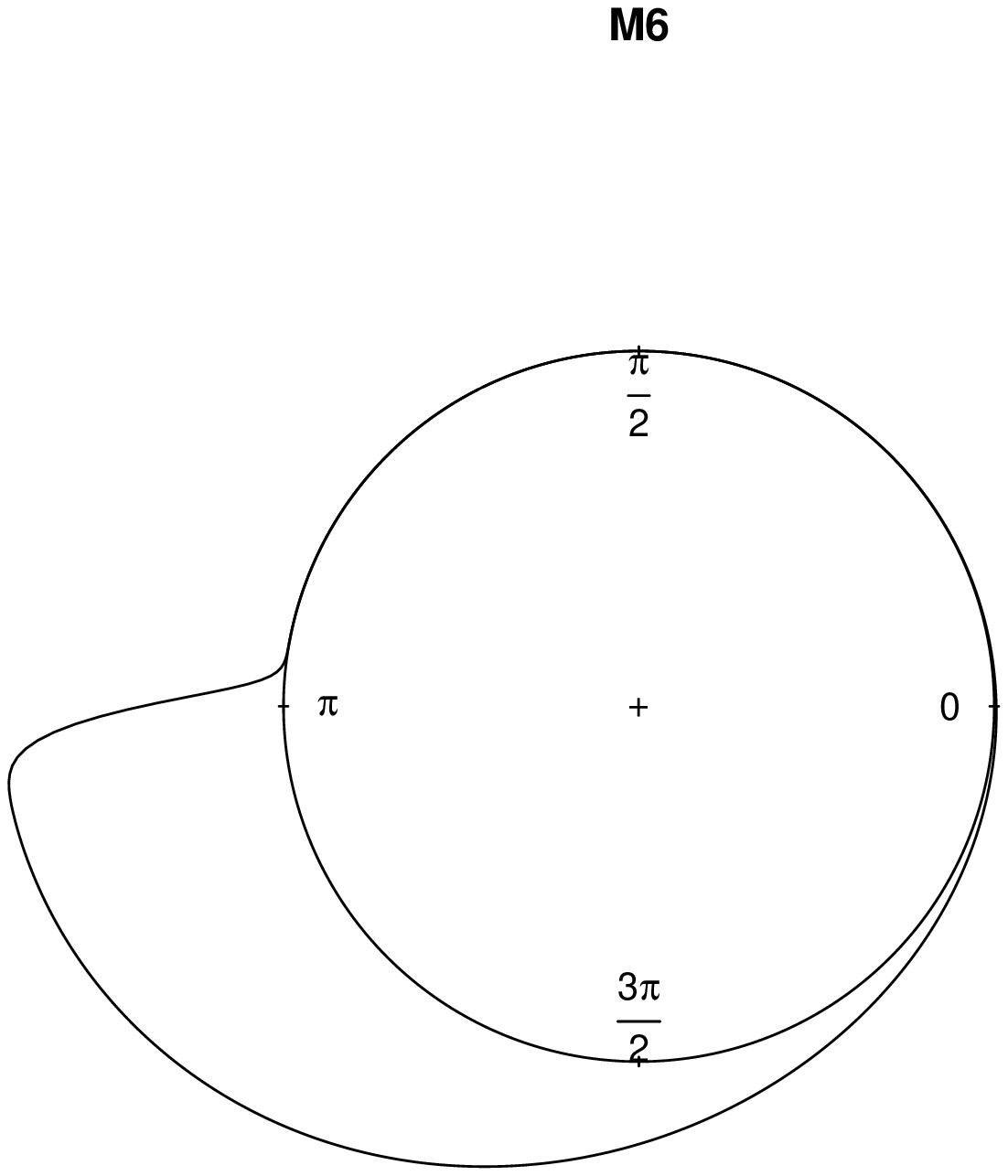}
\includegraphics[scale=0.22]{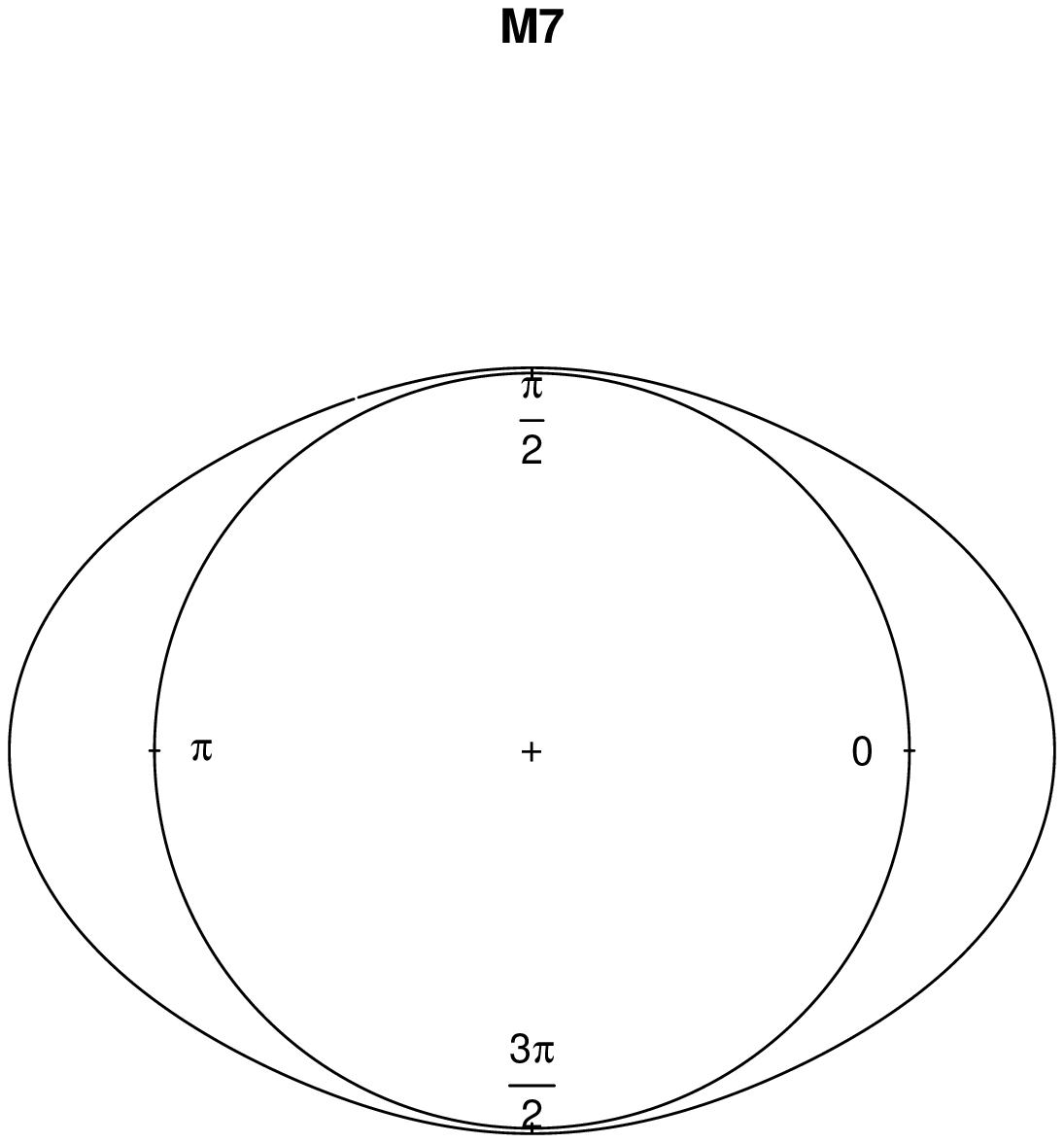}
\includegraphics[scale=0.22]{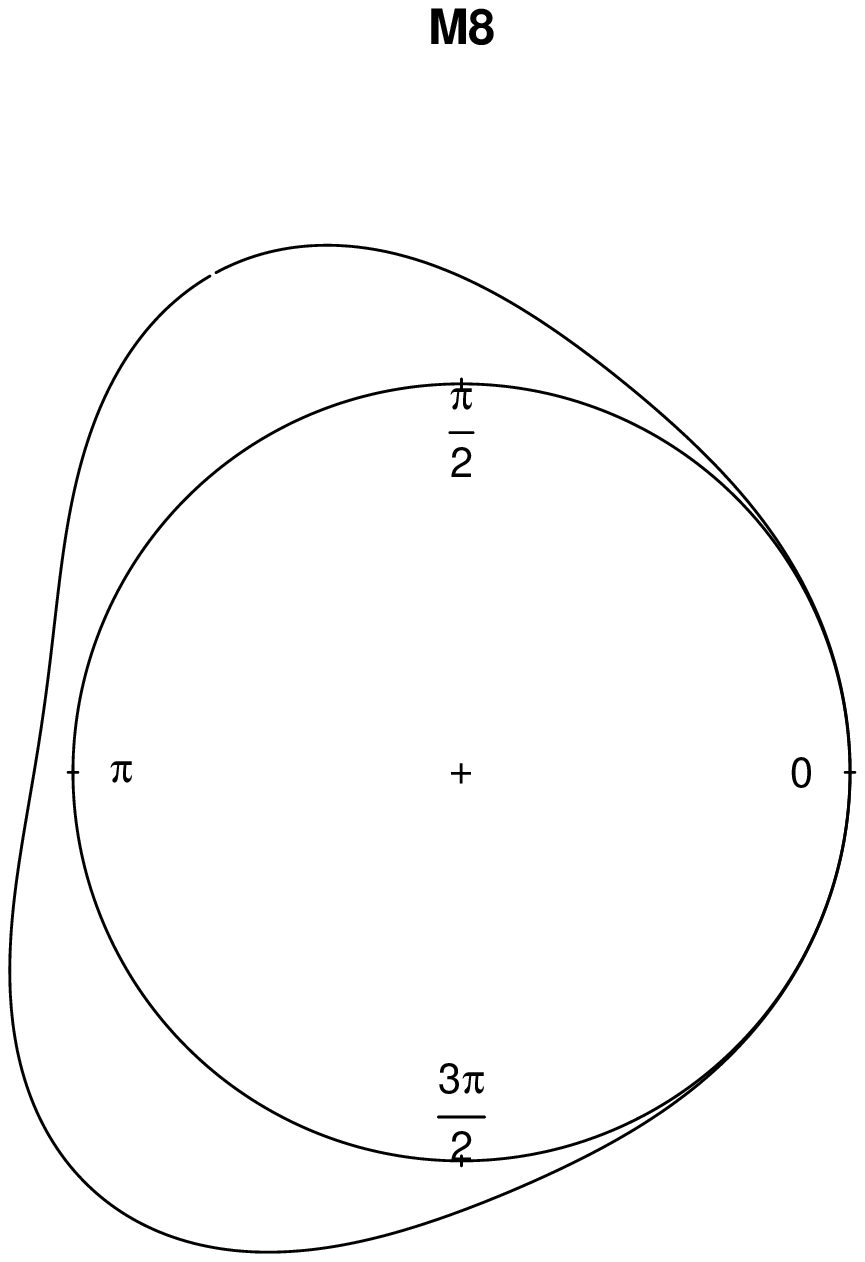}\\
\includegraphics[scale=0.22]{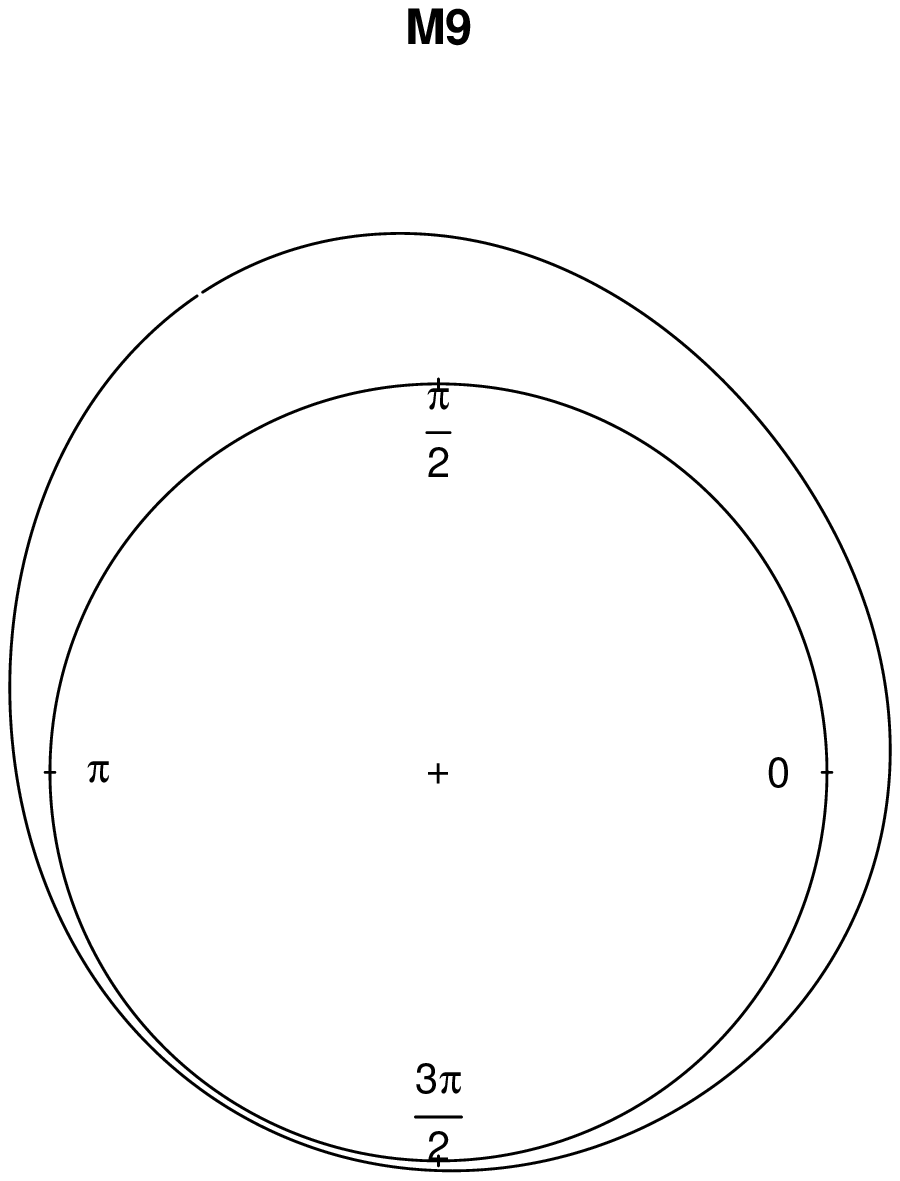}
\includegraphics[scale=0.22]{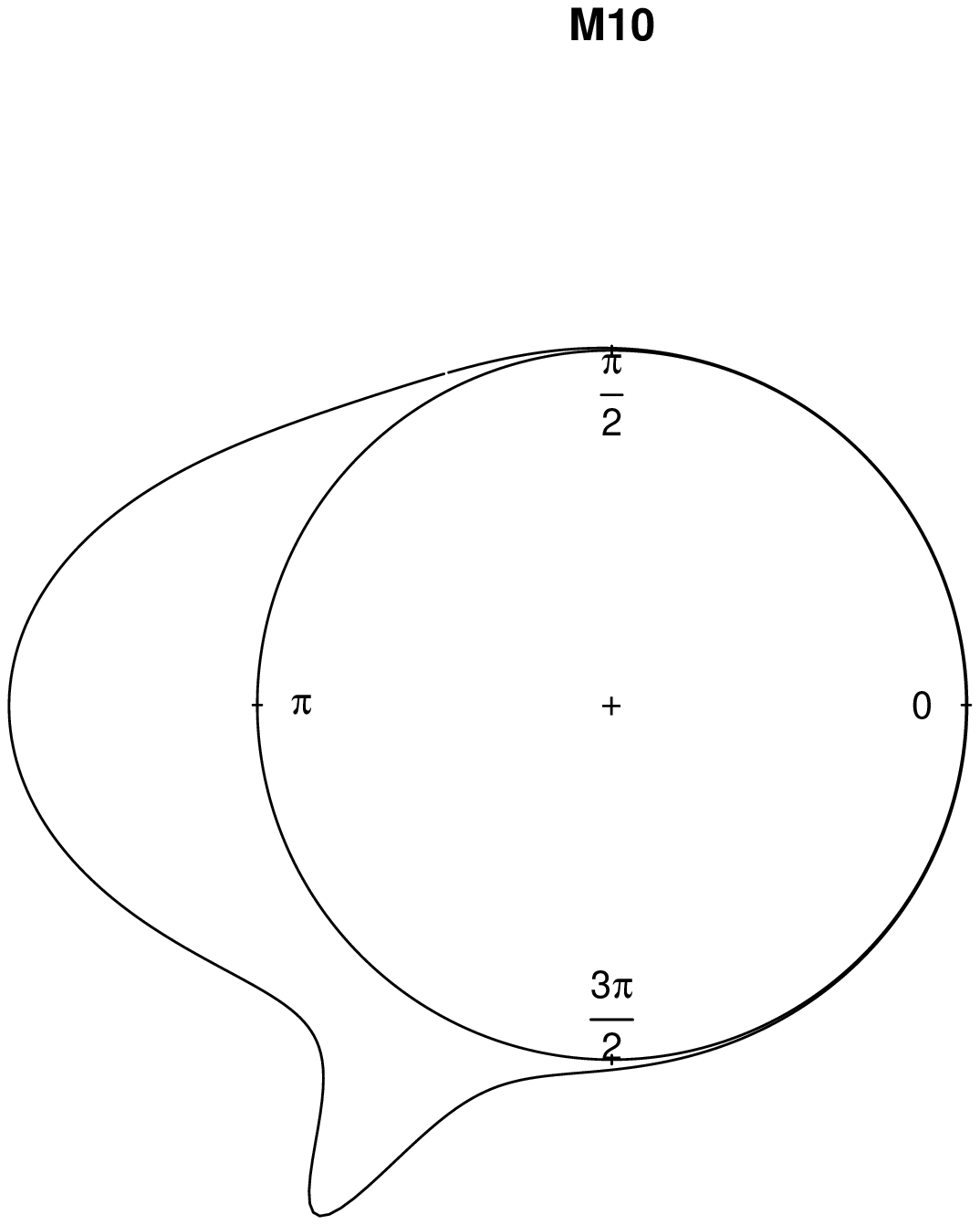}
\includegraphics[scale=0.22]{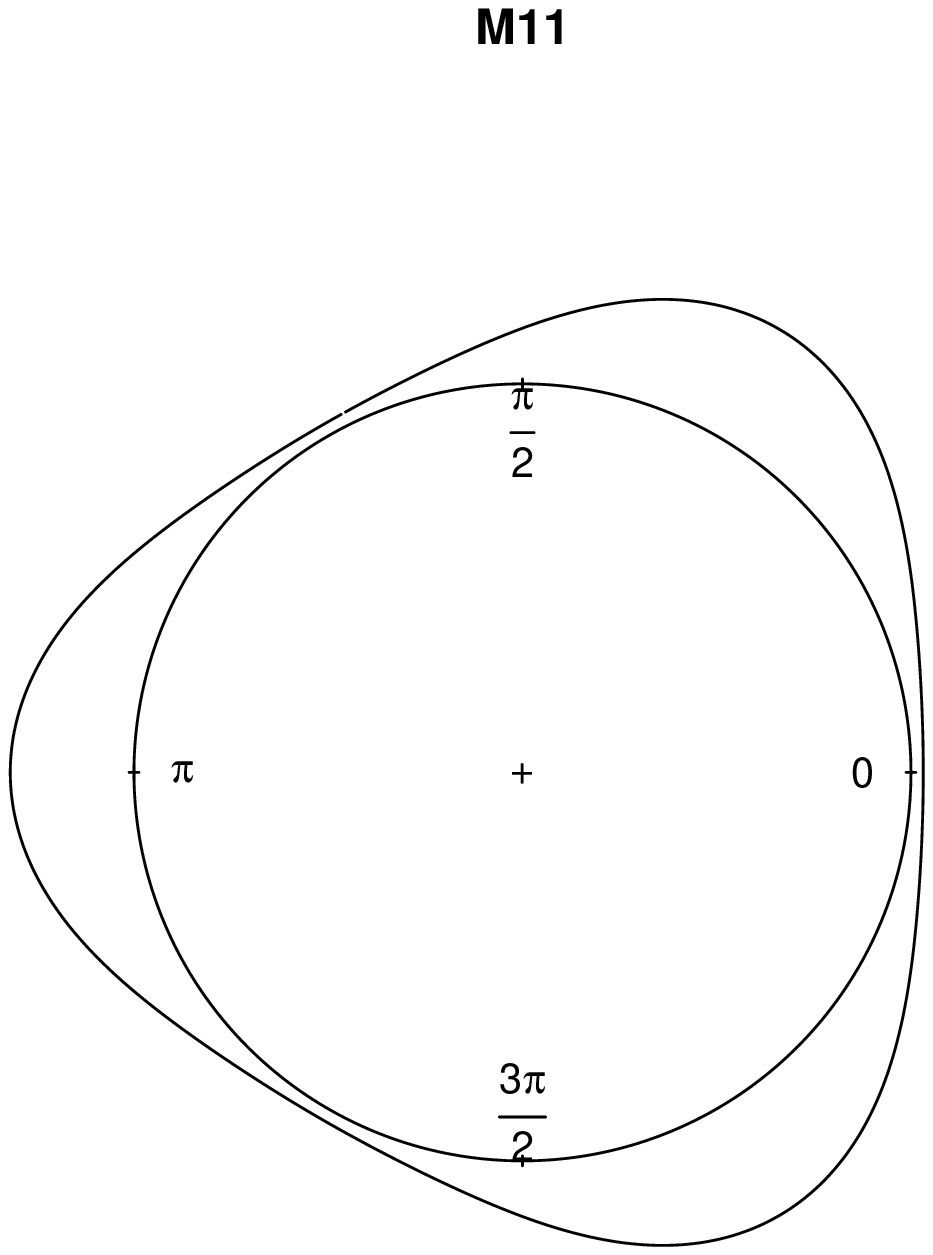}
\includegraphics[scale=0.22]{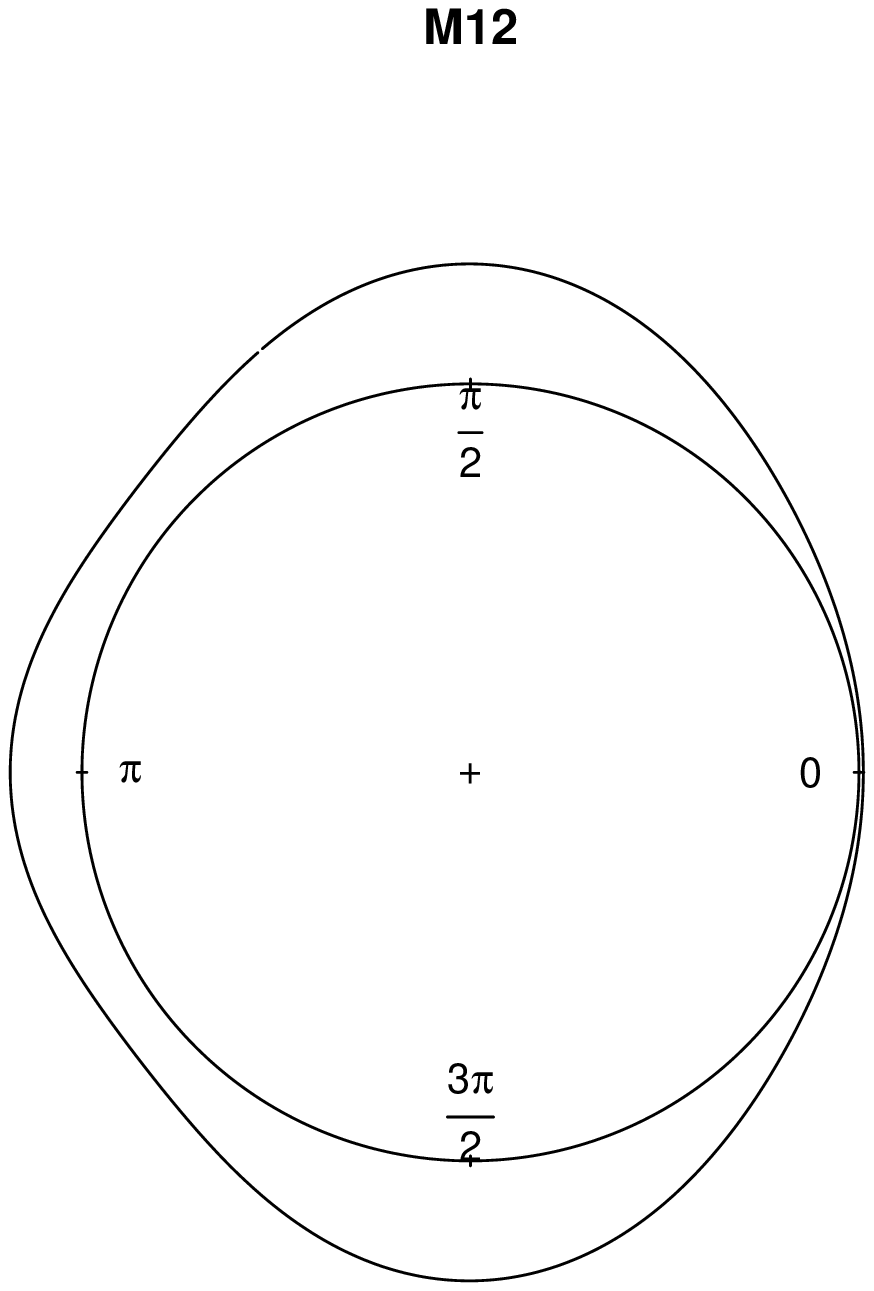}\\
\includegraphics[scale=0.22]{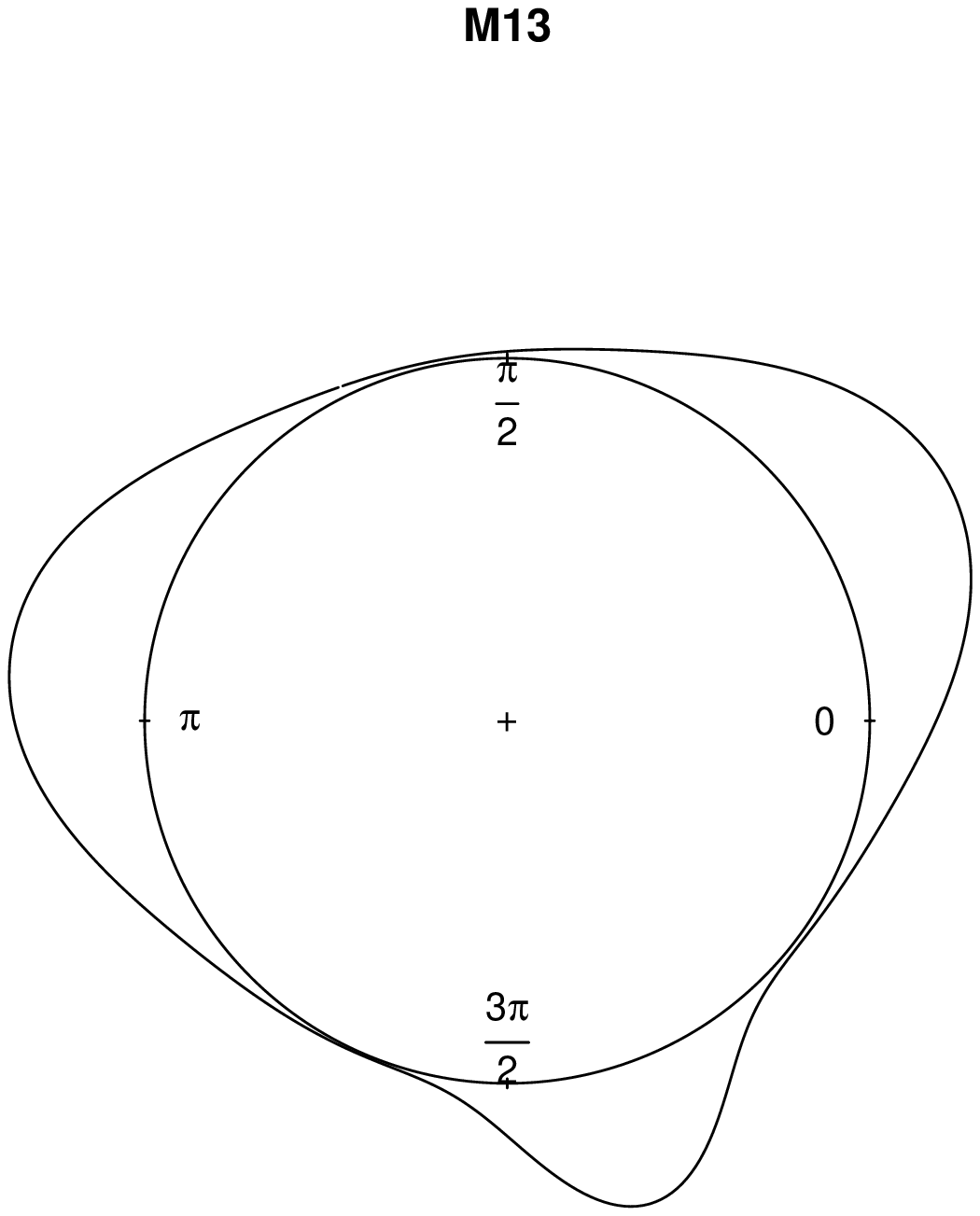}
\includegraphics[scale=0.22]{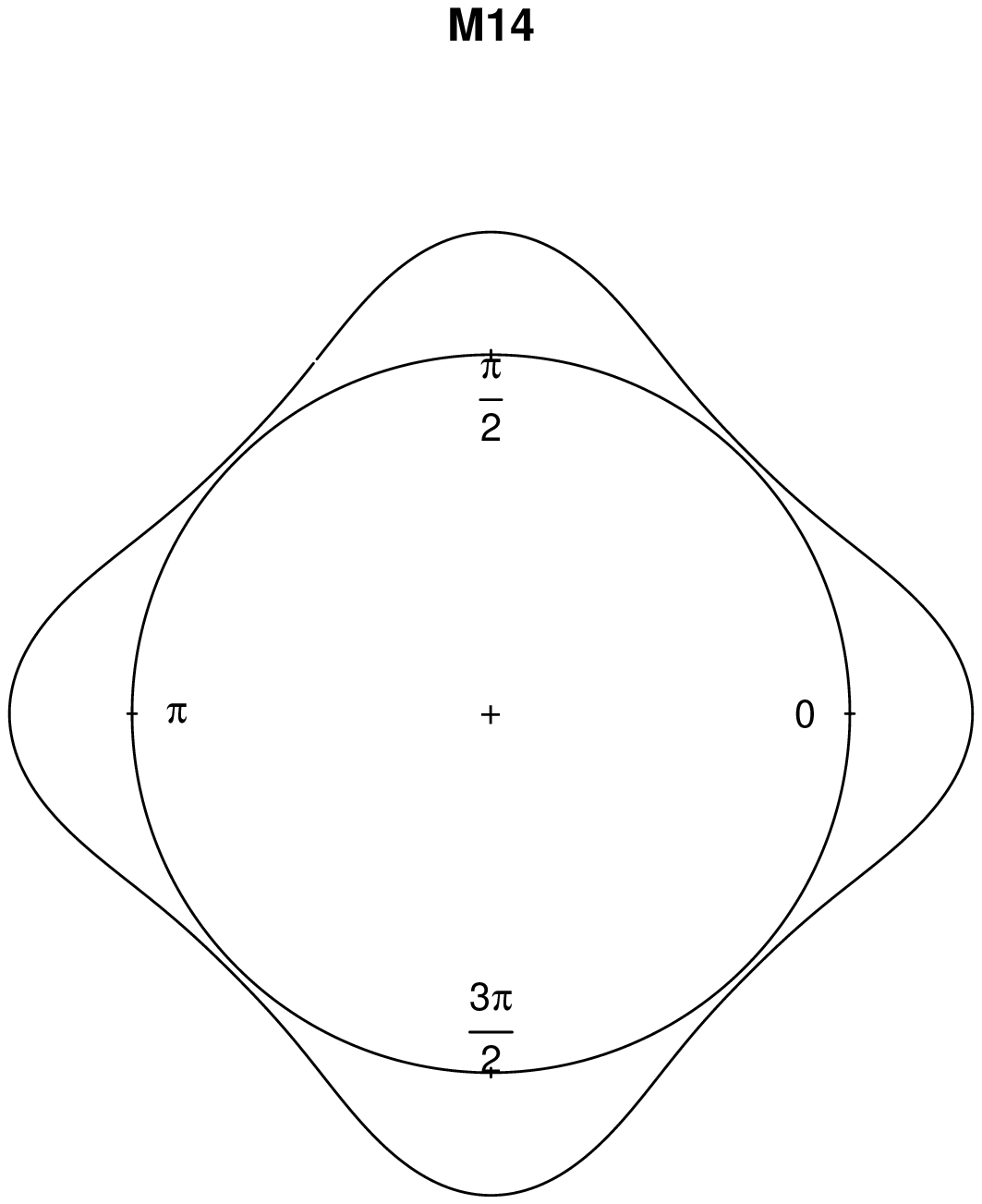}
\includegraphics[scale=0.22]{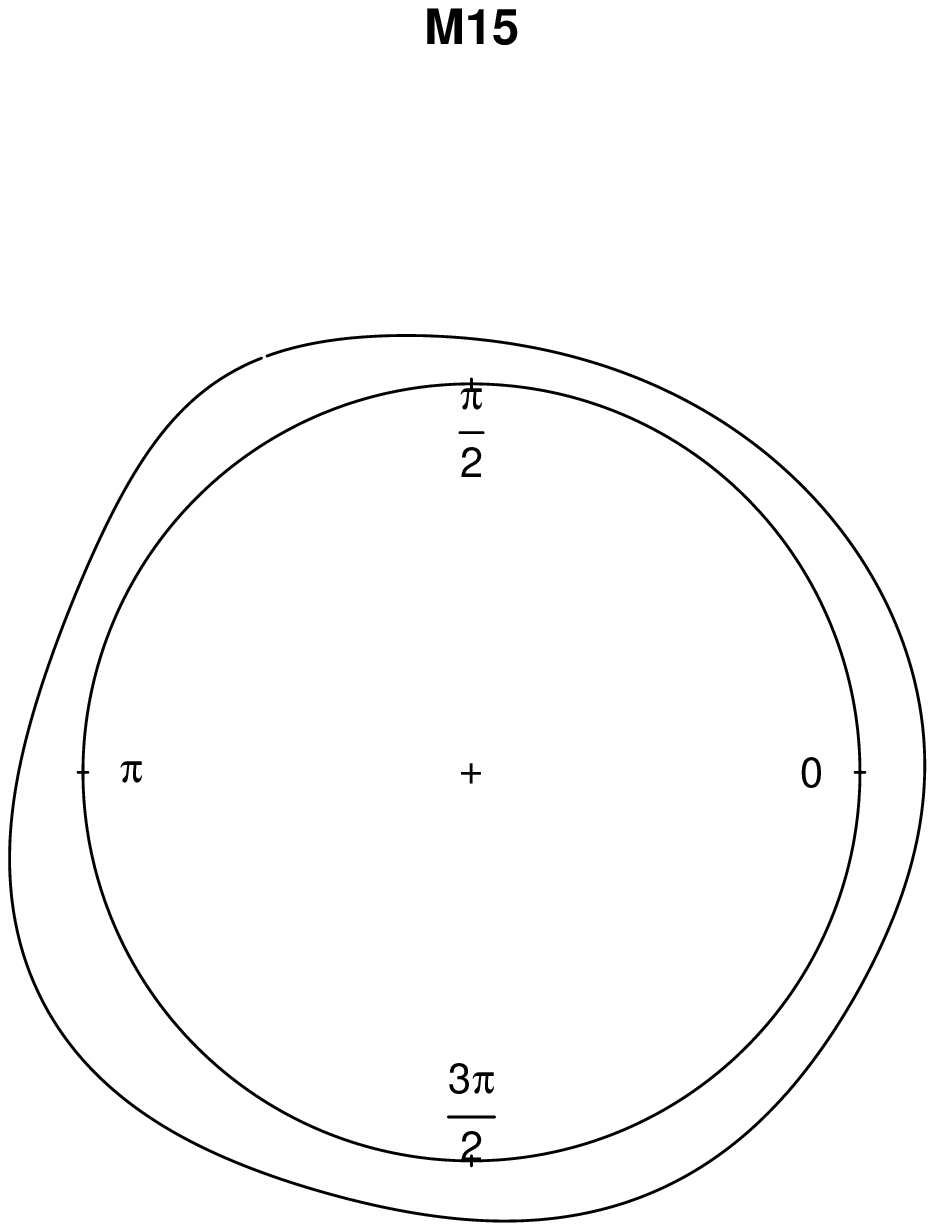}
\includegraphics[scale=0.22]{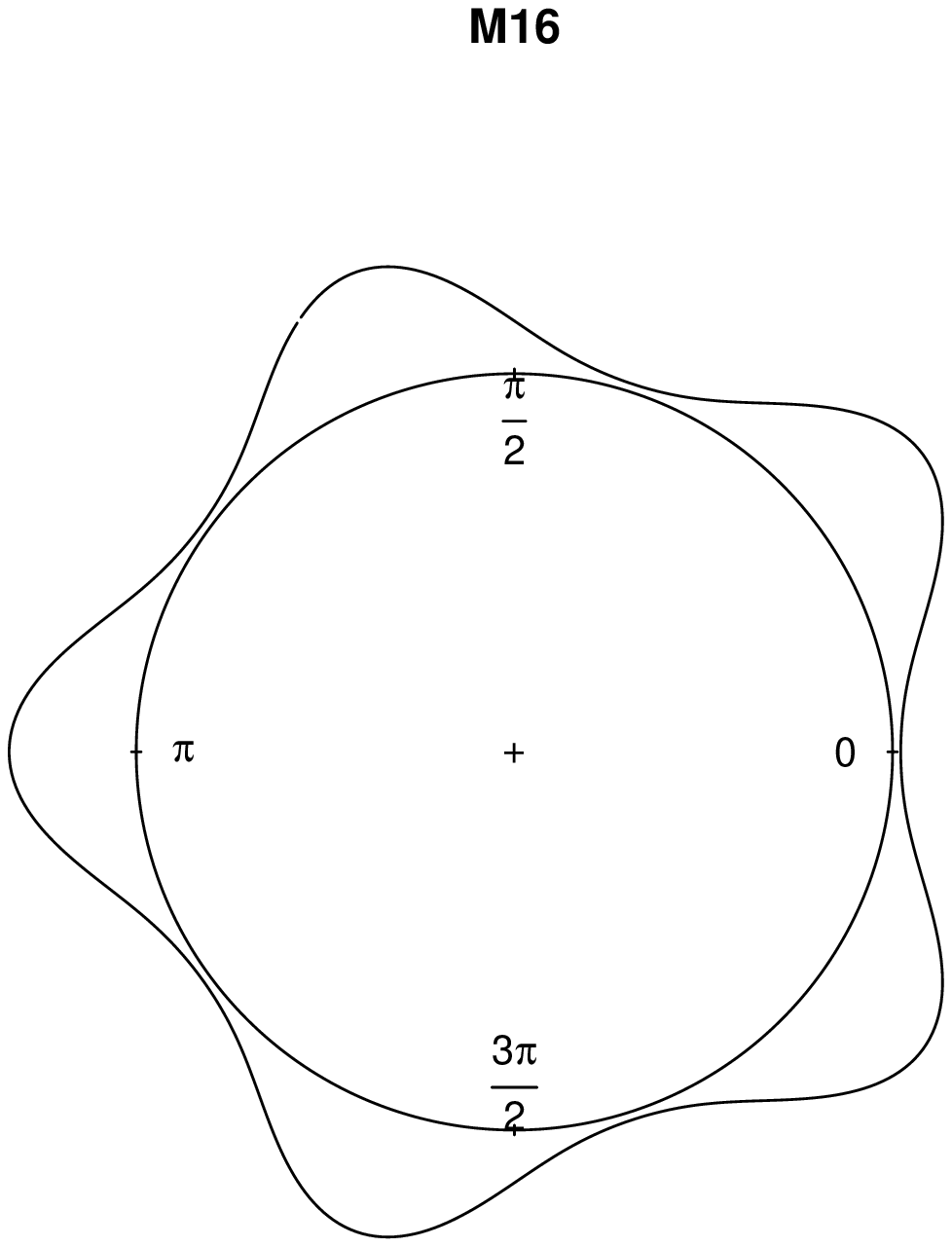}\\
\includegraphics[scale=0.22]{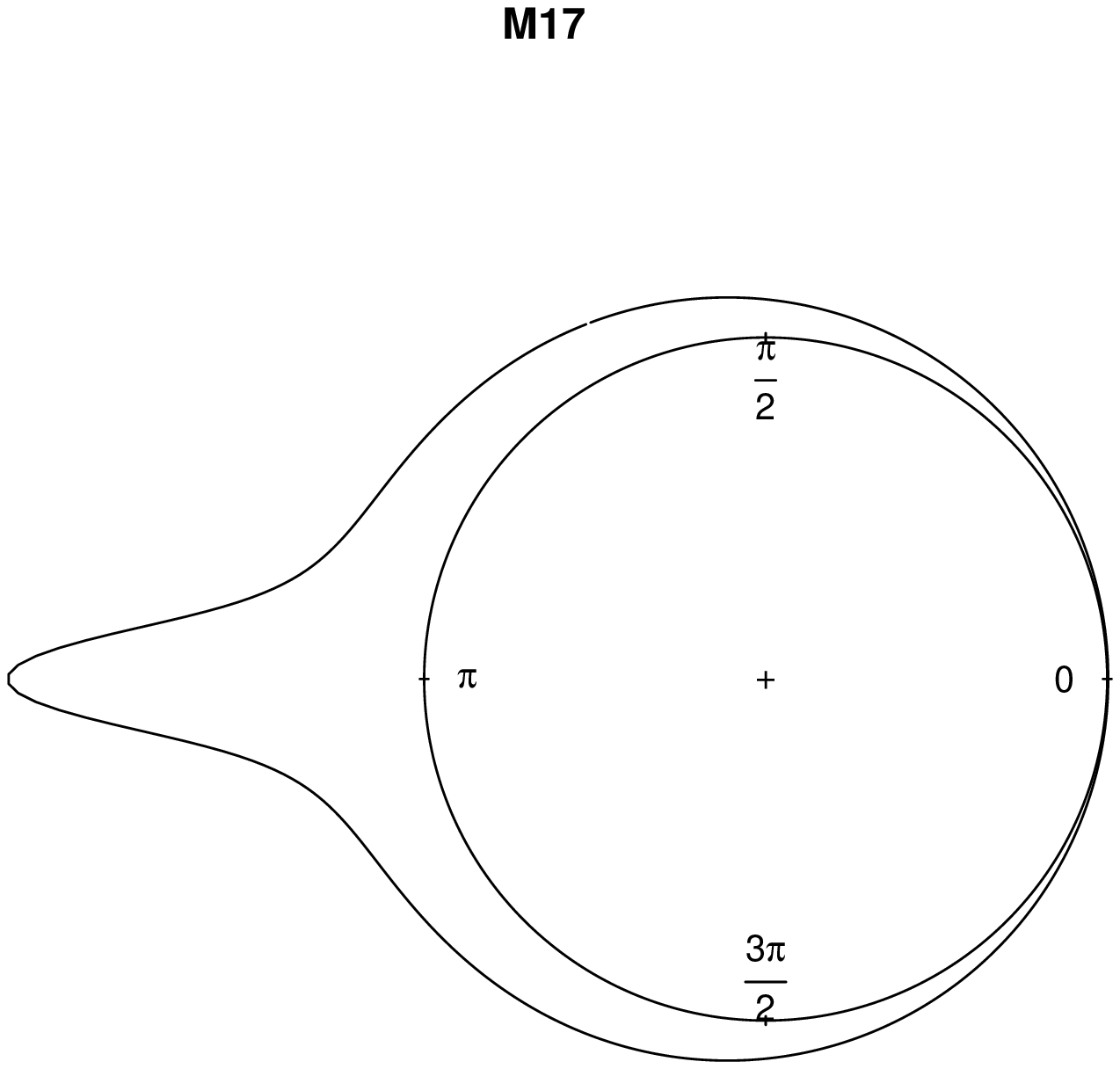}
\includegraphics[scale=0.22]{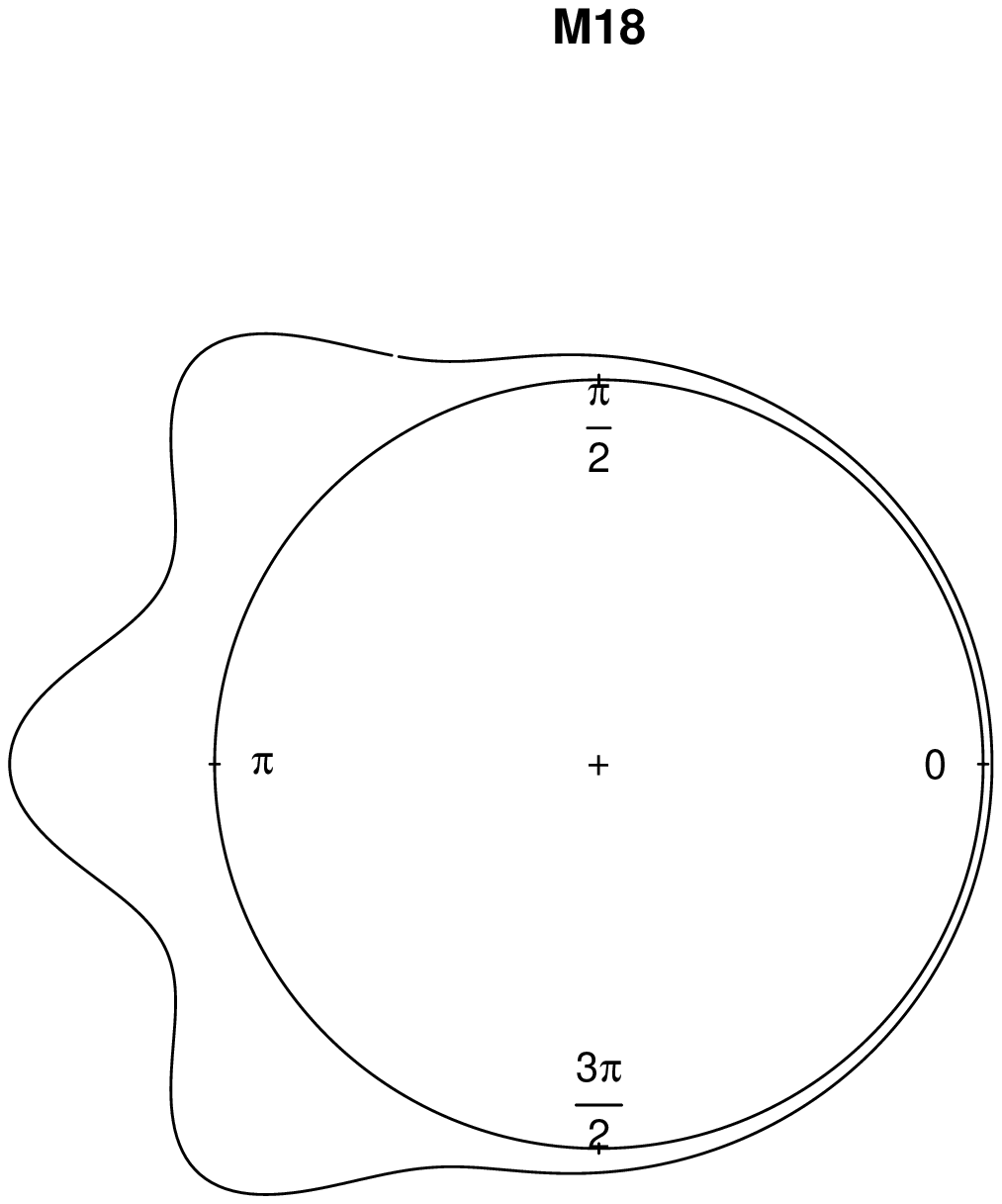}
\includegraphics[scale=0.22]{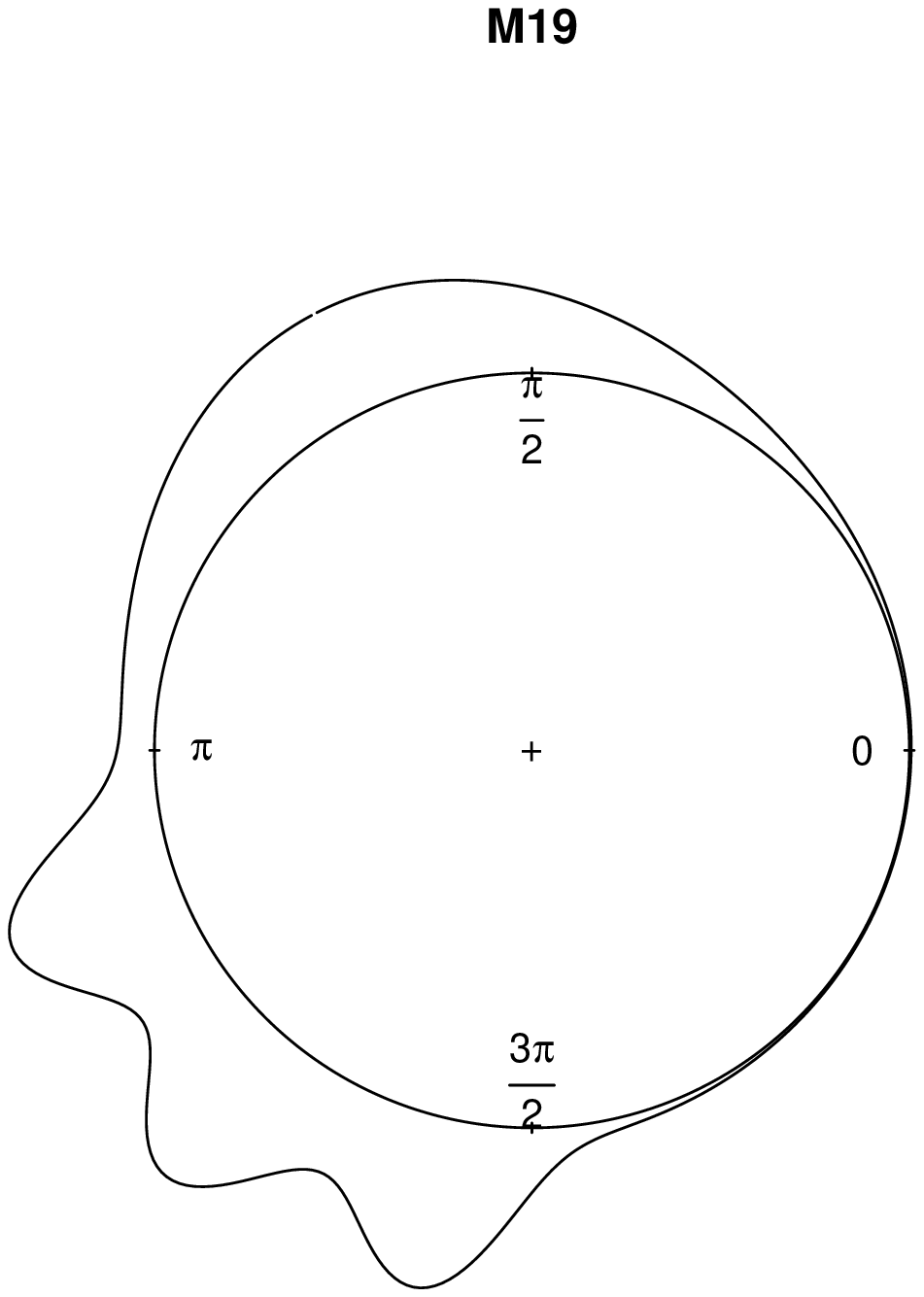}
\includegraphics[scale=0.22]{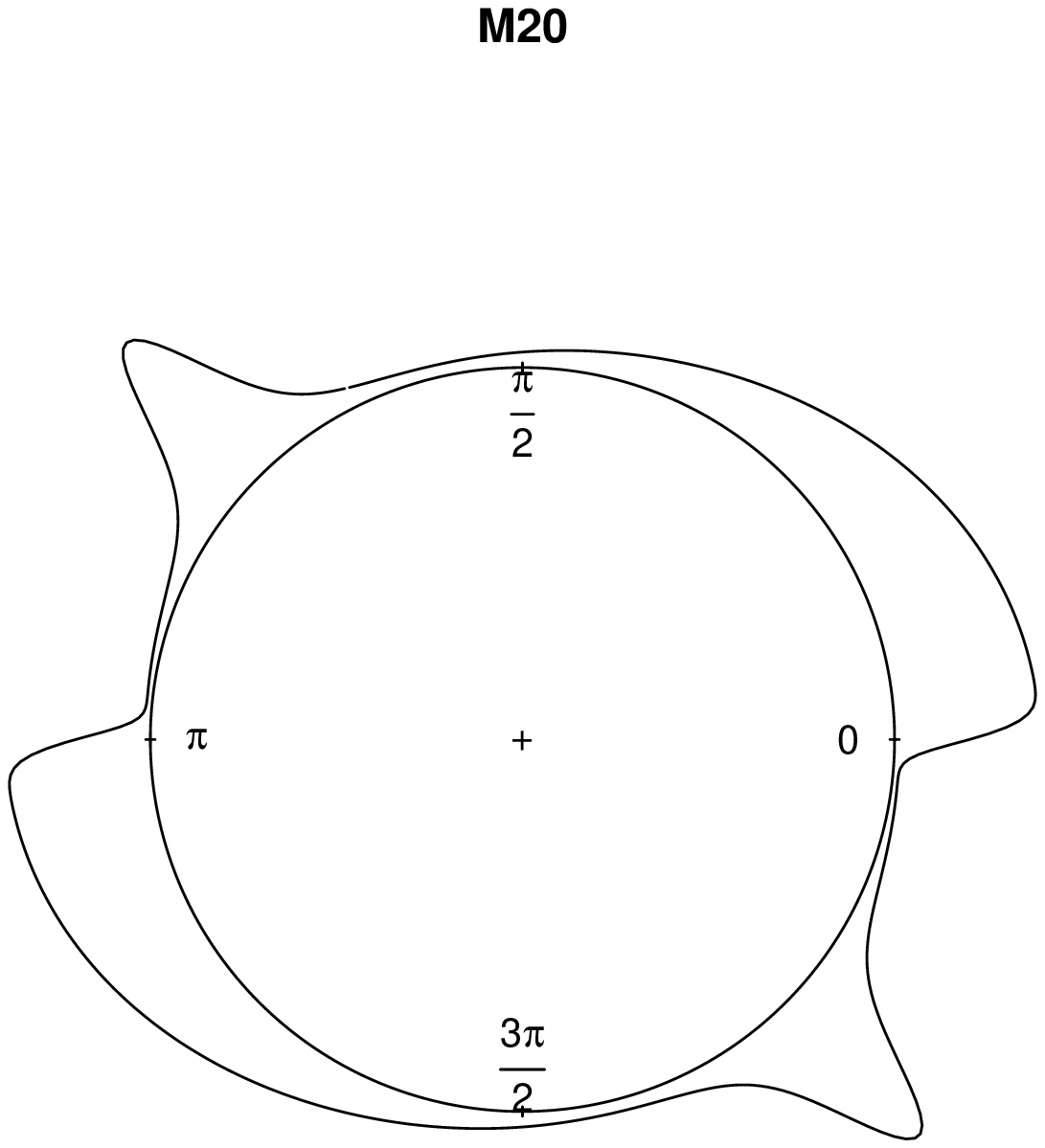}\\
\end{center}
\caption{Circular density models. M1-M6: simple models. M7-M10: two components models. M11-M16: models with three or more components. M17-M20: complex models.}
\label{distribution_models}
\end{figure}

\section{Simulation study}
\label{simulations}

The efectiveness of the new bandwidth selection method described in the previous section has been compared with the rule of thumb defined in (\ref{PI-selector}) and likelihood cross--validation (\ref{LCV}) through Monte Carlo experiments. A variety of circular distributions (von Mises, cardioid, various wrapped distributions and mixtures of them) displaying multimodality, skewness and/or peakedness have been tried (see Figure \ref{distribution_models} for plots and the Appendix for specific formulae). Technical details on these distribution models can be found in Jammalamadaka and SenGupta (2001), Mardia and Jupp (2000) and Pewsey (2000). For illustration purposes, the models have been classified in four groups, according to their complexity:

\textit{Simple models:} circular uniform (M1); von Mises (M2); wrapped Normal (M3); cardioid (M4); wrapped Cauchy (M5) and wrapped skew--Normal (M6).

\textit{Two components models:} von Mises mixtures (M7, M8 and M9); mixture of von Mises and wrapped Cauchy (M10).

\textit{Models with more than two components:} von Mises mixtures with three components (M11, M12 and M13); von Mises mixture with four components (M14); mixture of wrapped Cauchy, wrapped Normal, von Mises and wrapped skew--Normal (M15); von Mises mixture with five components (M16).

\textit{Other complex models:} mixture of cardioid and wrapped Cauchy (M17); mixture of von Mises (M18 and M19); mixture of two wrapped skew--Normal and two wrapped Cauchy (M20).

Note that \textit{Simple models} include unimodal models from von Mises distributions, with the circular uniform as a particular case. The wrapped Cauchy shows a highly peaked mode, whereas an asymmetric model is obtained with the wrapped skew--Normal, as shown in Pewsey (2006). The \textit{Two components models} collect different mixtures of two von Mises distributions (with antipodal modes and combining different weights and centers) and a mixture of a von Mises and a wrapped Cauchy, which results in a distribution with two modes with different concentrations. In \textit{Models with more than two components}, there are mixtures of three, four and five equally spaced and equally weighted von Mises distributions. Other situations with mild modes such as model M15 are also considered. Finally, \textit{Other complex models} are also included in the study. Although the distributions in this group are generated by mixtures of two or more models, the appearance may show a single mode, as in M17.

For each distribution model, $1000$ random samples of sizes $n=100,\, 250$ and $500$ were generated. Simulations were also obtained for sample size $1000$, with similar results to those corresponding to $500$ data, so they are ommitted.  In Tables \ref{Table_n100}, \ref{Table_n250} and \ref{Table_n500}, the average integrated squared errors, ISE ($ISE=\int (\hat f-f)^2$) of the circular kernel density estimator (\ref{kernel_circular}), considering different bandwidth selectors are shown. For each bandwidth selector, the average ISE over the 1000 replicates will be denoted, for the sake of simplicity, by $MISE(\nu)$. Specifically, the performance of the new plug--in rule $\hat\nu_{PI}$ will be compared with the rule of thumb $\hat\nu_{RT}$ and the cross--validation bandwidth $\hat\nu_{LCV}$. As a benchmark, the minimum average ISE has been computed for a broad grid of bandwidth parameters, denoted in the tables by $MISE(\nu_0)$.

In Section 2, a brief outline of the proposed algorithm for bandwidth selection has been given. Specific details in order to clarify its implementation in practice will be provided along this section. The simulations have been carried out in \emph{R} (see R Development Core Team, 2011), with self--programmed code for the kernel density estimator, the rule of thumb bandwidth proposed by Taylor (2008) the cross--validation bandwidth and the plug--in rule proposal. 

Step 1 in the algorithm requires the selection of the number of mixtures for the reference distribution. Note that the rule of thumb proposed in Taylor (2008) corresponds to $M=1$. We have tried the procedure with fixed $M$ in all the scenarios, obtaining $\hat\nu_{PI}^{M}$, for $M=2,3,4,5$ and observing that even with $M=2$, the plug--in rule gives better results than $M=1$. Just for illustrating our conclusions, the MISE values for $\hat\nu_{RT}$ and $\hat\nu_{PI}^{M=2}$ can be seen in Table \ref{Table_n100}, for $n=100$.

\begin{table}[h!]
\begin{center}
\scalebox{0.8}{
\begin{tabular}{c|c|ccc|c}
\hline
\hline
$n=100$  & $MISE(\nu_0)$  & $MISE(\hat\nu_{RT})$ & $MISE(\hat\nu_{PI})$ & $MISE(\hat\nu_{LCV})$&$MISE(\hat\nu_{PI}^{M=2})$   \\
\hline
\hline
M1  & 0.0000  & 0.0178 (0.0536) & 0.7434 (1.2332) & 0.3549 (0.6993) & 0.2953 (0.5396) \\
M2  & 0.5111  & 0.6532 (0.5100) & 1.0876 (1.2038) & 0.7016 (0.5963) & 0.7237 (0.7330) \\
M3  & 1.2218  & 1.2125 (0.8249) & 1.8720 (1.6935) & 1.4663 (1.1126) & 1.4571 (1.1753) \\
M4  & 0.4571  & 0.5188 (0.3672) & 1.1382 (1.2044) & 0.7508 (0.6843) & 0.6449 (0.5165) \\
M5  & 2.6790  & 8.4005 (2.7058) & 3.2661 (1.8693) & 6.6835 (3.0030) & 3.1759 (1.7627) \\
M6  & 2.3222  & 3.1492 (0.9402) & 3.3555 (1.8434) & 2.8453 (1.2634) & 2.6463 (1.2556) \\
\hline
M7  & 1.1144  & 10.5487 (0.3990) & 1.5229 (1.1190) & 1.3053 (0.7127) & 1.2135 (0.6425)\\
M8  & 1.2429  & 3.7140 (0.6896) & 1.6597 (1.2780) & 1.4247 (0.7842) & 1.3099 (0.7253)\\
M9  & 0.6713  & 0.7740 (0.5403) & 1.1062 (1.1424) & 0.7923 (0.5742) & 0.7664 (0.5186)\\
M10 & 2.3196  & 2.8927 (0.8219) & 3.2854 (1.7652) & 3.0658 (1.1273) & 2.8061 (1.2256) \\
\hline
M11 & 1.3439  & 6.4848 (0.0139) & 1.7230 (1.0791) & 1.5059 (0.6894) & 2.0585 (1.6312) \\
M12 & 1.0267  & 4.1352 (0.5260) & 1.5914 (1.1517) & 1.1802 (0.6284) & 1.1145 (0.5722)\\
M13 & 1.7347  & 10.8607 (0.1515) & 2.2178 (1.2374) & 1.8942 (0.8078) & 1.9368 (0.7983)\\
M14 & 1.8250  & 8.1836 (0.0470) & 2.1748 (1.0386) & 1.9584 (0.7770) & 7.1778 (1.8919)\\
M15 & 0.7012  & 0.7522 (0.0967) & 1.3558 (1.0468) & 0.9400 (0.5004) & 0.9040 (0.4104) \\
M16 & 2.2248  & 7.8368 (0.0607) & 2.4674 (0.9832) & 2.3189 (0.8273) & 7.5017 (1.1427) \\
\hline
M17 & 3.4773  & 7.8386 (1.1160) & 4.5339 (1.7917) & 5.5322 (1.8421) & 6.0295 (1.8351) \\
M18 & 2.1737  & 3.5950 (0.5300) & 3.1864 (1.4266) & 2.9698 (0.9083) & 3.2665 (0.8376) \\
M19 & 2.3469  & 3.8557 (0.6318) & 3.0182 (1.3658) & 2.5712 (0.7458) & 2.4593 (0.6747) \\
M20 & 3.2860  & 10.9618 (0.0540) & 4.0089 (1.5118) & 3.5078 (0.9101) & 6.4475 (1.3492) \\
\hline
\hline
\end{tabular}}
\caption{{\footnotesize Average integrated squared error for different bandwidth selectors, MISE ($\times 100$), and standard deviations ($\times 100$, in parentheses). Bandwidth selectors: $\hat\nu_{RT}$ (rule of thumb), $\hat\nu_{PI}$ (plug--in rule), $\hat\nu_{LCV}$ (likelihood cross--validation). $MISE(\nu_0)$: benchmark average integrated squared error. Sample size: $n=100$. Models M1--M20 distributed by complexity: M1--M6 (simple models); M7--M10 (two components models); M11--M16 (models with more than two components); M17--M20 (other complex models).}}
\label{Table_n100}
\end{center}
\end{table}
 
Small values of $M$ are suitable for simple models for any sample size (see Table \ref{Table_n100} for $M=2$ and $n=100$), and large values of $M$ are a good choice for complex models and moderate and large sample sizes. Hence, fixing the number of mixtures $M$ does not produce satisfactory results in all the simulation scenarios. The AIC criterion provides a data driven procedure for selecting $M$, so Step 1 in the algorithm is done as follows: AIC is computed for mixtures of $M= 2, 3, 4, 5$ von Mises distributions and the selected number of mixtures $M$ for the reference distribution is the one minimizing the AIC.

For sample size $n=100$ (see Table \ref{Table_n100}), the plug--in rule $\hat\nu_{PI}$ is competitive with the other bandwidth selectors. The AIC criterion tends to select a large value for $M$, which may be damaging in some simple models compared with the results for $\hat\nu_{PI}^{M=2}$. Therefore, one should not try AIC with a large number of mixtures in the reference distribution. Besides, for small sample sizes, it may not be realistic to attempt to estimate too complicated models (see $MISE(\nu_0)$ in Table \ref{Table_n100} for M17 to M20).

\begin{table}[h!]
\begin{center}
\scalebox{0.8}{
\begin{tabular}{c|c|ccc}
\hline
\hline
$n=250$ &  $MISE(\nu_0)$ & $MISE(\hat\nu_{RT})$ & $MISE(\hat\nu_{PI})$ & $MISE(\hat\nu_{LCV})$   \\ 
\hline
\hline
M1  & 0.0000 & 0.0037 (0.0119) & 0.1499 (0.2848) & 0.1321 (0.2628) \\
M2  & 0.2568 & 0.3201 (0.2211) & 0.3499 (0.2866) & 0.3610 (0.2891) \\
M3  & 0.6072 & 0.6510 (0.4357) & 0.7954 (0.6808) & 0.7517 (0.5525) \\
M4  & 0.2418 & 0.2485 (0.1556) & 0.3948 (0.3337) & 0.3521 (0.2845) \\
M5  & 1.4101 & 5.8159 (1.5250) & 1.6012 (0.8717) & 2.9313 (1.3962) \\
M6  & 1.3422 & 2.1665 (0.5032) & 1.6544 (0.7329) & 1.5842 (0.6379) \\
\hline
M7  & 0.5762 & 10.6753 (0.1786) & 0.5986 (0.3400) & 0.5976 (0.2917) \\
M8  & 0.6466 & 2.3765 (0.4021) & 0.6961 (0.4185) & 0.7105 (0.3841) \\
M9  & 0.3473 & 0.4162 (0.2467) & 0.4196 (0.3116) & 0.4171 (0.2704) \\
M10 & 1.3545 & 2.0187 (0.4325) & 1.5816 (0.6363) & 2.0316 (0.6941) \\
\hline
M11 & 0.6766 & 6.4797 (0.0016) & 0.7358 (0.3678) & 0.7368 (0.3278) \\
M12 & 0.5232 & 3.7831 (0.4806) & 0.6108 (0.3249) & 0.6010 (0.3062) \\
M13 & 0.8890 & 10.8954 (0.0922) & 0.9456 (0.3761) & 0.9234 (0.3503) \\
M14 & 0.9105 & 8.1691 (0.0080) & 0.9690 (0.3705) & 0.9675 (0.3608) \\
M15 & 0.4301 & 0.7285 (0.0840) & 0.6027 (0.2769) & 0.5381 (0.2311) \\
M16 & 1.1141 & 7.8224 (0.0117) & 1.1355 (0.3753) & 1.1473 (0.3857) \\
\hline
M17 & 1.8929 &  6.6517 (0.7443) & 2.2035 (0.8596) & 3.2325 (1.2213) \\
M18 & 1.1325 & 2.9559 (0.2953) & 1.3480 (0.6393) & 1.4273 (0.5726) \\
M19 & 1.3048 & 3.0017 (0.3424) & 1.5400 (0.5080) & 1.5813 (0.4046) \\
M20 & 1.8126 & 10.9744 (0.0396) & 1.9511 (0.5298) & 1.9224 (0.4740) \\
\hline
\hline
\end{tabular}}
\caption{{\footnotesize Average integrated squared error for different bandwidth selectors, MISE ($\times 100$), and standard deviations ($\times 100$, in parentheses). Bandwidth selectors: $\hat\nu_{RT}$ (rule of thumb), $\hat\nu_{PI}$ (plug--in rule), $\hat\nu_{LCV}$ (likelihood cross--validation). $MISE(\nu_0)$: benchmark average integrated squared error. Sample size: $n=250$. Models M1--M20 distributed by complexity: M1--M6 (simple models); M7--M10 (two components models); M11--M16 (models with more than two components); M17--M20 (other complex models). }}
\label{Table_n250}
\end{center}
\end{table}

Nevertheless, including a complex reference distribution, i.e. large $M$, is reasonable for large enough datasets. The AIC criterion succeeds in selecting a suitable $M$ as can be seen in all the considered scenarios. For moderate and large sample sizes ($n=250,\,500$), results with the AIC selection equal or even outperform the best $\hat\nu_{PI}^{M}$. The strength of the new proposal can be seen in Tables \ref{Table_n250} and \ref{Table_n500}. 

In more detail, for \emph{Simple models} and $n=250,\,500$, the three bandwidth selectors show a similar behaviour in models M1 to M4. As expected, the rule of thumb outperforms the other selectors in model M1, which corresponds with the circular uniform. However, the behaviour shown by the rule of thumb in models M5 and M6 is quite poor, compared with the plug--in selector which is the best in these cases. Likelihood cross--validation is worse than plug--in, although better than the rule of thumb. Note that M5 is the wrapped Cauchy distribution and M6 is the wrapped skew--Normal, confirming the adecquate performance of the plug--in rule for estimating highly peaked and asymmetric distributions.

\begin{table}
\begin{center}
\scalebox{0.8}{
\begin{tabular}{c|c|ccc}
\hline
\hline
$n=500$  & $MISE(\nu_0)$ & $MISE(\hat\nu_{RT})$ & $MISE(\hat\nu_{PI})$ & $MISE(\hat\nu_{LCV})$   \\ 
\hline
\hline
M1  & 0.0000  & 0.0011 (0.0032) & 0.0488 (0.1039) & 0.0804 (0.1698) \\
M2  & 0.1622  & 0.1924 (0.1320) & 0.1854 (0.1250) & 0.2104 (0.1508) \\
M3  & 0.3837  & 0.3760 (0.2353) & 0.4051 (0.2821) & 0.4183 (0.2646) \\
M4  & 0.1439  & 0.1490 (0.0922) & 0.2067 (0.1459) & 0.2158 (0.1498) \\
M5  & 0.8544  & 4.1746 (0.9683) & 0.9264 (0.4794) & 1.4937 (0.7707) \\
M6  & 0.8244  & 1.6077 (0.3009) & 0.9377 (0.3601) & 0.9764 (0.3651) \\
\hline
M7  & 0.3259  & 10.7146 (0.1268) & 0.3346 (0.1546) & 0.3528 (0.1674) \\
M8  & 0.3767  & 1.6509 (0.2643) & 0.3925 (0.1849) & 0.4174 (0.1916) \\
M9  & 0.2117  & 0.2602 (0.1543) & 0.2346 (0.1454) & 0.2521 (0.1572) \\
M10 & 0.8562  & 1.5684 (0.2866) & 0.9472 (0.3397) & 1.4131 (0.4573) \\
\hline
M11 & 0.3932  & 6.4796 (0.0025) & 0.4097 (0.1666) & 0.4304 (0.1839) \\
M12 & 0.3169  & 3.3673 (0.4107) & 0.3326 (0.1470) & 0.3436 (0.1550) \\
M13 & 0.5333  &10.8987 (0.0816) & 0.5527 (0.2051) & 0.5567 (0.2071) \\
M14 & 0.5365  & 8.1673 (0.0025) & 0.5585 (0.1987) & 0.5688 (0.2028) \\
M15 & 0.2685  & 0.7146 (0.0739) & 0.3383 (0.1707) & 0.3212 (0.1491) \\
M16 & 0.6434  & 7.8200 (0.0032) & 0.6548 (0.2019) & 0.6683 (0.2142) \\
\hline
M17 & 1.1769  & 5.7398 (0.5669) & 1.3342 (0.5357) & 2.0593 (0.7907) \\
M18 & 0.6654  & 2.5307 (0.2169) & 0.7299 (0.2851) & 0.8107 (0.3219) \\
M19 & 0.7930  & 2.4971 (0.2077) & 0.8974 (0.2992) & 0.9961 (0.2854) \\
M20 & 1.1043  & 10.9869 (0.0231) & 1.1711 (0.2836) & 1.1696 (0.2810) \\
\hline
\hline
\end{tabular}}
\caption{{\footnotesize Average integrated squared error for different bandwidth selectors, MISE ($\times 100$), and standard deviations ($\times 100$, in parentheses). Bandwidth selectors: $\hat\nu_{RT}$ (rule of thumb), $\hat\nu_{PI}$ (plug--in rule), $\hat\nu_{LCV}$ (likelihood cross--validation). $MISE(\nu_0)$: benchmark average integrated squared error. Sample size: $n=500$. Models M1--M20 distributed by complexity: M1--M6 (simple models); M7--M10 (two components models); M11--M16 (models with more than two components); M17--M20 (other complex models).}}
\label{Table_n500}
\end{center}
\end{table}

In the \emph{Two components models}, the performance of $\hat\nu_{RT}$ is extremely poor for model M7 (antipodal modes), and is also far from satisfactorily for models M8 and M10. The plug--in rule $\hat\nu_{PI}$ provides good results for all the models in this group (compared with the optimal MISE, $MISE(\nu_0)$), whereas $\hat\nu_{LCV}$ seems to be a competitor except for model M10. 

For the next group of models, \emph{More than two components}, the rule of thumb seems not consistent (except for model M15, which is  \emph{almost} flat). The plug--in rule and the likelihood cross--validation bandwidth behave similarly. Finally, for \emph{Other complex models} (except for M20), the plug--in rule outperforms the other selectors.

Summarizing, for $n=250$ and after testing for significant differences between the MISE values, we can conclude that the plug--in rule is the best option in models M5, M10, M17, M18 and M19. For the other models (except for flat or almost falt models M1, M4 and M15), the plug--in rule and the likelihood cross--validation selector behave similarly. For $n=500$, the plug--in rule provides the best results in all the scenarios, being the likelihood cross--validation selector competitive in M1, M3, M4, M12, M13, M14, M15, M16 and M20. Thus, for moderate and large sample sizes, the proposed plug--in selector is the best for most models, and in general, it is always a good alternative.

\section{Illustration with real data}

In this section, the circular kernel density estimator is used to analyze two real data examples. Both of them are classical examples of asymmetric and bimodal distribution, respectively, regarding azimuths of cross--beds in a river and animal orientation behaviour.

\textbf{Example 1. Cross--beds azimuths.} A classical dataset that shows an asymmetric distribution corresponds to azimuths of cross--beds in the Kamthi river. Originally analyzed by SenGupta and Rao (1966) and included in Table 1.5 in Mardia (1972), the dataset collects 580 azimuths of layers lying oblique to the principal accumulation surface along the river, being these layers known as cross--beds.

\begin{figure}[h!]
\begin{center}
\includegraphics[width=6cm]{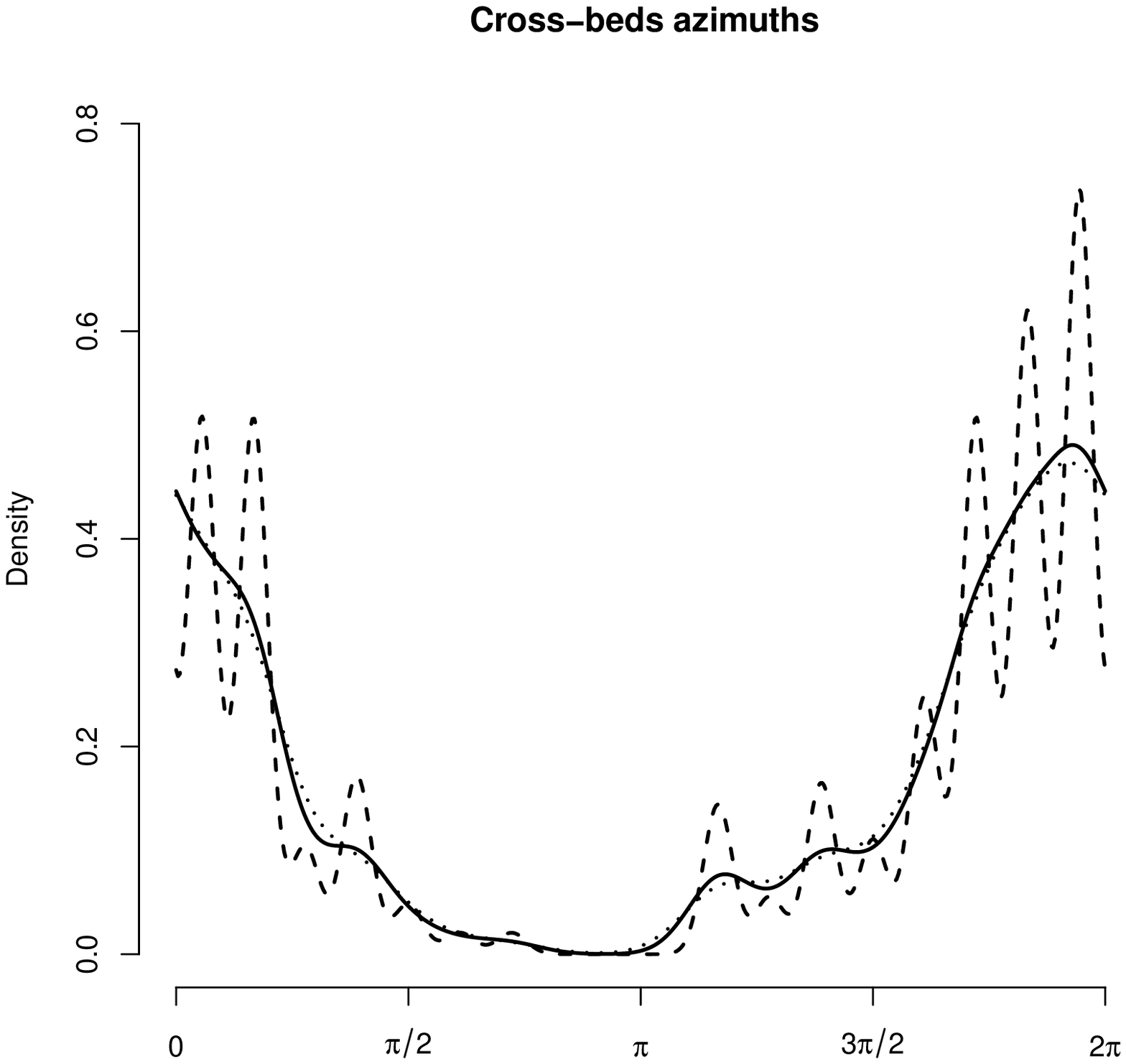}
\includegraphics[width=6cm]{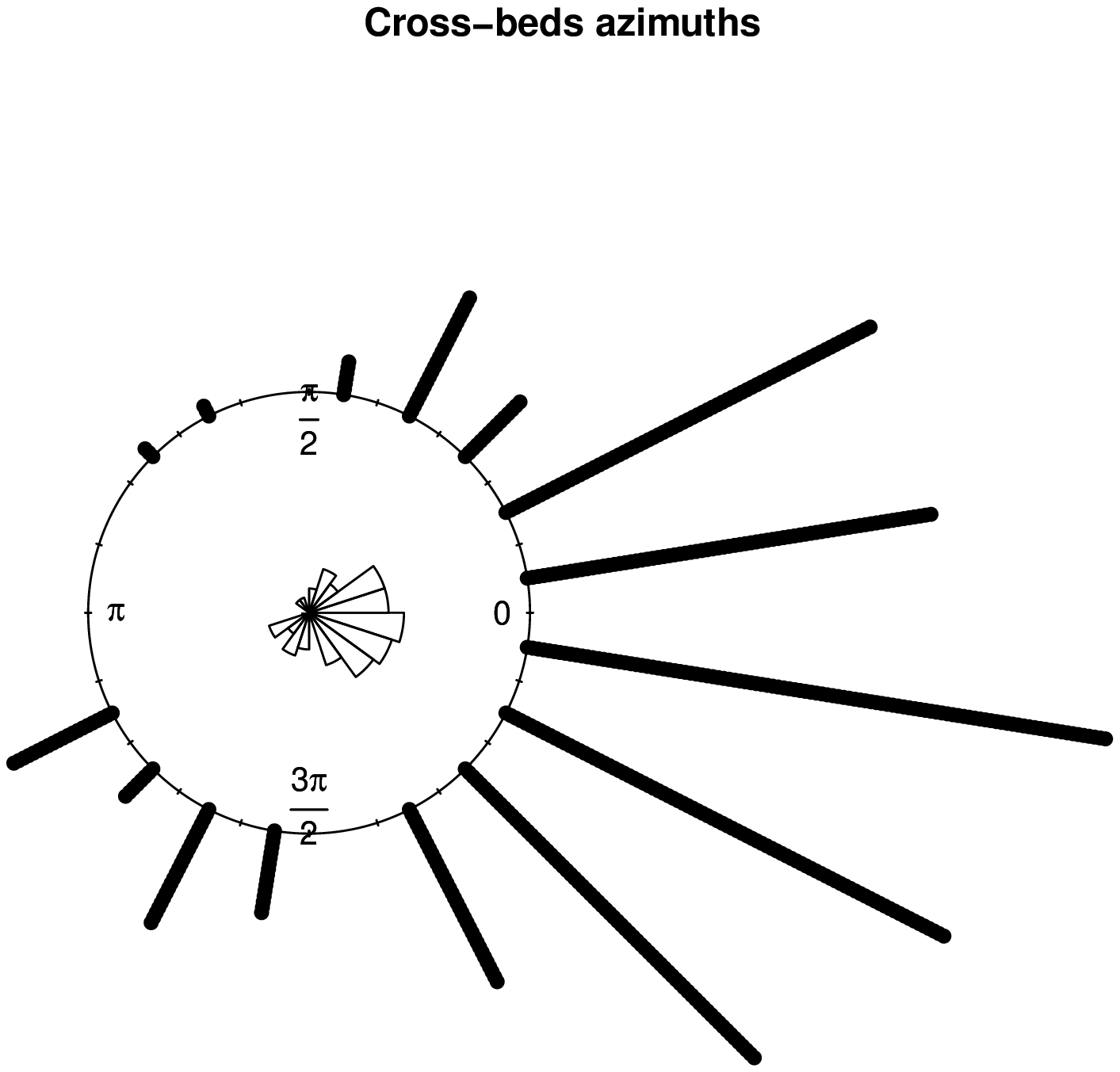}
\caption{Circular kernel density estimators for the azimuths (left panel) and rose diagram (right panel). Solid line: plug--in selector, $\hat\nu_{PI}$. Dashed line: likelihood cross--validation bandwidth, $\hat\nu_{LCV}$. Dotted line: rule of thumb, $\hat\nu_{RT}$. }
\label{azi}
\end{center}
\end{figure}

A circular kernel density estimation has been computed for this dataset, considering three different bandwidth selectors: rule of thumb, $\hat\nu_{RT}$, likelihood cross--validation, $\hat\nu_{LCV}$ and plug--in rule, $\hat\nu_{PI}$. In Figure \ref{azi} (left panel), it can be seen that the estimators with the rule of thumb and the plug--in bandwidths perform similarly, fitting a unimodal distribution with negative (anticlockwise) asymmetry. However, the likelihood cross--validation criterion provides a too large smoothing parameter, resulting in an undersmoothed fitted density. In this case, the number of selected mixtures by AIC was $M=2$.

\begin{figure}[h!]
\begin{center}
\includegraphics[width=6cm]{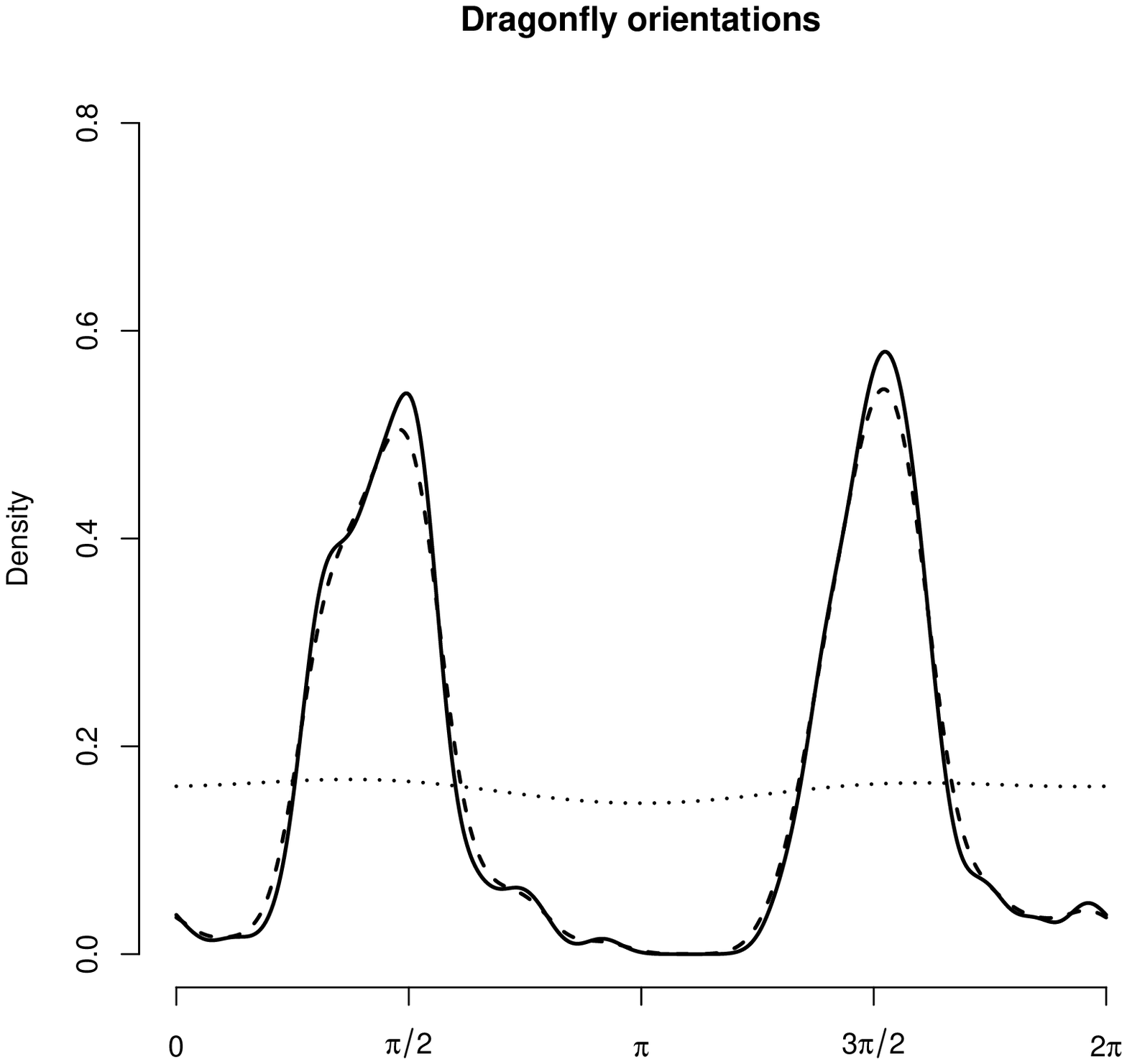}
\includegraphics[width=6cm]{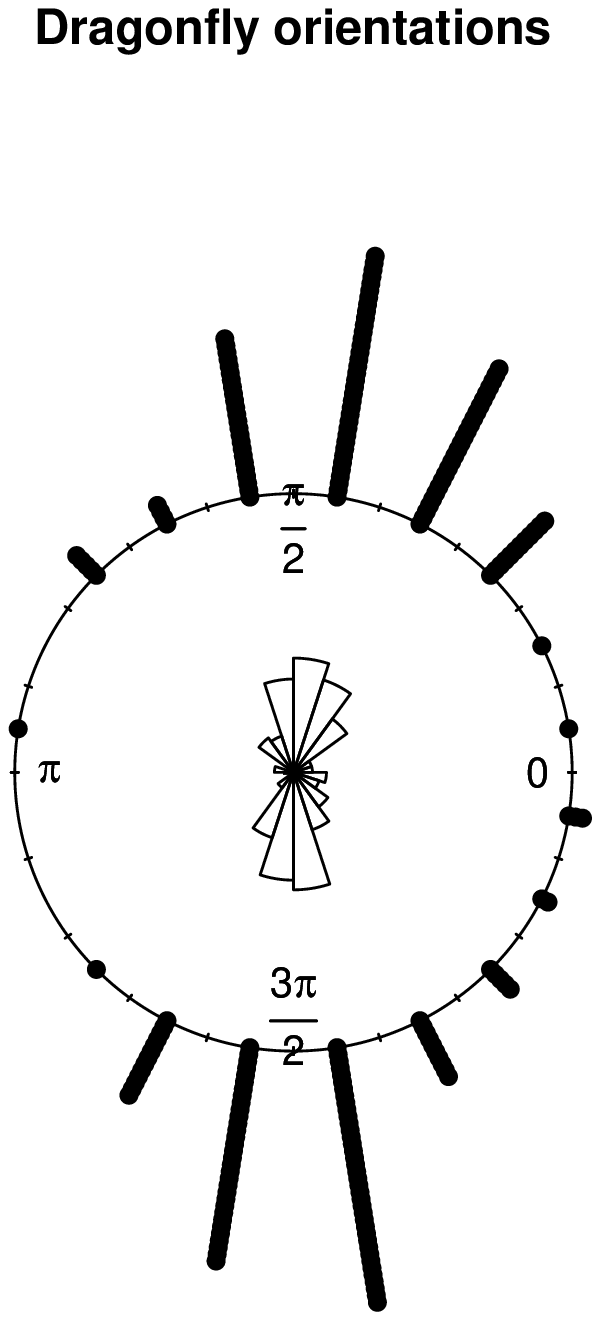}
\caption{Circular kernel density estimators for the dragonflies (left panel) and rose diagram (right panel) data. Solid line: plug--in selector, $\hat\nu_{PI}$. Dashed line: likelihood cross--validation bandwidth, $\hat\nu_{LCV}$. Dotted line: rule of thumb, $\hat\nu_{RT}$. }
\label{drag}
\end{center}
\end{figure}

\textbf{Example 2. Dragonflies orientation.} Circular data also arises in animal orientation studies. Among the examples presented in Batschelet (1981), we consider the orientation of 214 dragonflies with respect to the sun's azimuth. As it can be seen already from the circular plot in Figure \ref{drag} (right panel), this is a clear example of bimodal circular distribution. This dataset was also studied by Pewsey (2004), who applied a test for circular reflective symmetry. 
In a situation like this one (opposite modes), the rule of thumb behaves quite poorly, as can be seen in Figure \ref{drag} (left panel). Likelihood cross--validation and plug--in selectors provide similar fitted curves, showing two modes. The AIC criterion selects $M=4$ mixtures for the reference distribution.

\section*{Final comments}
The proposed procedure behaves satisfactorily for all the simulation scenarios and the real data examples, at a moderate computational cost in comparison with the likelihood cross--validation selector. For instance, for the data examples, it took 3.82 seconds to compute the likelihood cross--validation bandwidth, whereas the plug--in rule selector was obtained in 1.17 seconds (with R code running in a regular laptop). 

In practice, computational problems may disable the AIC output. These difficulties may appear in the implementation of the EM algorithm (which is available in the R package movMF) and/or from the numerical approximation of the integral in Step 2, which may not be finite. In this situation, the number of mixtures for the reference distribution is chosen as the one that provides the minimum valid AIC. Just when no results can be obtained for the different values of $M$,  the rule of thumb is chosen. It should be noticed that in our simulation study, this situation only occurred for model M3 (for $n=100$, 9 out of 1000 samples needed $M=1$, for $n=250$, 3 out of 1000 samples), M6 (for $n=100$, 2 out of 1000 samples) and M10 (for $n=100$, 37 out of 1000 samples; for $n=250$, 3 out of 1000 samples; for $n=500$, 1 out of 1000 samples). It does not seem to be an issue for large sample size.

As commented in Section \ref{kernel_section}, and from what is seen in the results for model M7, one of the problems of the rule of thumb in the presence of antipodal modes is that it tends to provide uniform estimates for the circular density, which corresponds to a null concentration parameter in the von Mises family. A natural question arises: what would happen if a different parametric family, not including the uniform distribution, is used as a reference? We have also checked by simulations, considering the same models as the ones presented here, that setting a wrapped Cauchy in the minimization of the AMISE error in (\ref{AMISE}) provides better results than the rule of thumb, but far from the new plug--in rule proposal.

\section*{Acknowledgements}
The authors want to acknowledge Prof. Arthur Pewsey for facilitating data examples and Prof. Jos\'e E. Chac\'on, for the fruitful comments and discussions. This work has been supported by Project MTM2008--03010 from the Spanish Ministry of Science and Innovation.




\newpage

\section*{Appendix}
The models included in Section \ref{simulations} for the simulation experiments are described in this Appendix (see Figure \ref{distribution_models}). Models from M1 to M20 were classified according to their complexity in four groups. However, for clarifying how they are obtained, they will be introduced regarding the parametric family or families used for their construction.

{\bf Circular uniform.} The circular uniform distribution is described in Section 2.2.1 of Jammalamadaka and SenGupta (2001) and Section 3.5.3 of Mardia and Jupp (2000). Model M1 is a circular uniform. It can be seen as a particular case of the von Mises distribution (see below) with null concentration parameter.

{\bf Von Mises distribution.} The von Mises distribution, $vM(\mu,\kappa)$, is a symmetric unimodal distribution characterized by a mean direction $\mu$ and concentration parameter $\kappa$. See Section 2.2.4 of Jammalamadaka and SenGupta (2001) and Section 3.5.4 of Mardia and Jupp (2000) for details. Model M2 is $vM(\pi,1)$. A variety of mixtures of von Mises distributions have been also considered. Specifically, models M7 ($1/2\cdot vM(0,4)+1/2\cdot vM(\pi,4)$), M8 ($1/2\cdot vM(2,5)+1/2\cdot vM(4,5)$) and M9 ($1/4\cdot vM(0,2)+3/4\cdot vM(\pi/\sqrt{3},2)$) are mixtures of von Mises with two components. Three mixtures of three von Mises have been also tried in models M11 ($1/3\cdot vM(\pi/3,6)+1/3 \cdot vM(\pi,6)+1/3 \cdot vM(5\pi/3,6)$), M12 ($2/5\cdot vM(\pi/2,4)+1/5 \cdot vM(\pi,5)+2/5\cdot vM(3\pi/2,4)$) and M13 ($2/5\cdot vM(0.5,6)+2/5 \cdot vM(3,6)+1/5\cdot vM(5,24)$). Equally weighted mixtures with four and five symmetrically distributed modes have been constructed in models M14 (mixture of $vM(0,12)$, $vM(\pi/2,12)$, $vM(\pi,12)$ and $vM(3\pi/2,12)$, with weigths equal to $1/4$) and M16 (mixture of $vM(\pi/5,18)$, $vM(3\pi/5,18)$, $vM(\pi,18)$, $vM(7\pi/5,18)$, $vM(9\pi/5,18)$, with weights equal to $1/5$). Other mixtures of four von Mises have been used in order to obtain complex models. More precisely, M18 and M19 correspond to $1/2\cdot vM(\pi,1)+1/6\cdot vM(\pi-0.8,30)+1/6\cdot vM(\pi,30)+1/6\cdot vM(\pi+0.8,30)$ and $4/9\cdot vM(2,3)+5/36\cdot vM(4,3)+5/36\cdot vM(3.5,50)+5/36\cdot vM(4,50)+5/36\cdot vM(4.5,50)$, respectively.

{\bf Cardioid distribution.} The cardioid distribution, $Cardioid(\mu,\rho)$, is unimodal and symmetric around the mean direction $\mu$ and with concentration parameter $\rho$. This distribution is the perturbation of the uniform density by a cosine funtion. See Section 2.2.2 of Jammalamadaka and SenGupta (2001) and Section 3.5.5 of Mardia and Jupp (2000). In the simulation study, M4 is a cardiod $Cardioid(0,0.5)$

{\bf Wrapped Normal distribution.} The wrapped Normal distribution $WN(\mu,\rho)$ is obtained by wrapping the $N(\mu,\sigma^2)$ distribution onto the circle, where $\rho=e^{-\sigma^2/2}$. It is unimodal and symmetric about its mode $\mu$. See Section 2.2.6 of Jammalamadaka and SenGupta (2001) and Section 3.5.7 of Mardia and Jupp (2000). Model M3 is $WN(0,0.9)$

{\bf Wrapped Cauchy distribution.} The wrapped Cauchy distribution, $WC(\mu,\rho)$, is obtained by wrapping the Cauchy distribution on the real line. The $WC(\mu,\rho)$ is unimodal and symmetric about $\mu$ and with concentration parameter $\rho$. See Section 2.2.7 of Jammalamadaka and SenGupta (2001) and Section 3.5.7 of Mardia and Jupp (2000). In the simulation study, M5 is $WC(0,0.8)$. In addition, M10 is a mixture of a von Mises distribution and a wrapped Cauchy ($4/5\cdot vM(\pi,5)+1/5\cdot WC(4\pi/3,0.9)$) and M17 is a mixture of a cardioid and a wrapped Cauchy ($2/3\cdot Cardioid(\pi,0.5)+1/3\cdot WC(\pi,0.9)$).

{\bf Wrapped skew--Normal distribution.} The wrapped skew--Normal distribution, $WSN(\xi,\eta,\lambda)$ is a skewed distribution characterized by a location parameter $\xi$, a scale parameter $\eta$ and a skewness parameter $\lambda$ (see Pewsey, 2000 for further details). The wrapped skew--Normal has been used in model M6, which is a $WSN(0,1,20)$, and for obtaining two complex models: M15, a mixture of a wrapped Cauchy, a wrapped Normal, a von Mises and a wrapped skew--Normal ($3/10\cdot WC(\pi-1,0.6)+ 1/4 \cdot WN(\pi+0.5,0.9)+1/4\cdot vM(\pi+2,3)+1/5\cdot WSN(6,1,3)$), and model M20, a mixture of two wrapped skew--Normal and two wrapped Cauchy ($1/3\cdot WSN(0,0.7,20)+1/3\cdot WSN(\pi,0.7,20)+1/6\cdot WC(3\pi/4,0.9)+1/6\cdot WC(7\pi/4,0.9)$).

\end{document}